\documentclass[fleqn,usenatbib]{mnras}

\usepackage{newtxtext,newtxmath,graphicx}

\usepackage[T1]{fontenc}

\DeclareRobustCommand{\VAN}[3]{#2}
\let\VANthebibliography\thebibliography
\def\thebibliography{\DeclareRobustCommand{\VAN}[3]{##3}\VANthebibliography}

\usepackage{graphicx}	
\usepackage{amsmath}	
\usepackage{CJKutf8}        
\usepackage{xcolor}     
\usepackage{xspace}
\usepackage{longtable}
\usepackage{hyperref}

\newcommand{\gaia}{\textit{Gaia}\xspace}

\title[Outskirts of classical dSphs with {\it Gaia} EDR3]{Stellar proper motions in the outskirts of classical dwarf spheroidal galaxies with {\it Gaia} EDR3}

\author[Qi et al.]{
Yuewen Qi(\begin{CJK*}{UTF8}{gbsn}齐\end{CJK*}\begin{CJK*}{UTF8}{bsmi}玥雯\end{CJK*})$^{1,2}$\thanks{E-mail: yuewenq@tamu.edu},
Paul Zivick$^{1,2}$, Andrew B. Pace$^{3}$, Alexander H.~Riley$^{1,2}$, and Louis E.~Strigari$^{1,2}$
\\
$^{1}$Mitchell Institute for Fundamental Physics and Astronomy, Texas A\&M University, College Station, TX 77843, USA \\
$^{2}$Department of Physics and Astronomy, Texas A\&M University, College Station, TX 77843, USA \\
$^{3}$ McWilliams Center for Cosmology, Carnegie Mellon University, 5000 Forbes Ave, Pittsburgh, PA 15213, USA
}

\date{Accepted XXX. Received YYY; in original form ZZZ}

\pubyear{2021}

\begin{document}
\label{firstpage}
\pagerange{\pageref{firstpage}--\pageref{lastpage}}
\maketitle

\begin{abstract}
We use Gaia EDR3 data to identify stars associated with six classical dwarf spheroidals (Draco, Ursa Minor, Sextans, Sculptor, Fornax, Carina) at their outermost radii, beyond their nominal King stellar limiting radius. For all of the dSphs examined, we find radial velocity matches with stars residing beyond the King limiting radius and with $> 50\%$ astrometric probability (four in Draco, two in Ursa Minor, eight in Sextans, two in Sculptor, twelve in Fornax, and five in Carina), indicating that these stars are associated with their respective dwarf spheroidals (dSphs) at high probability. We compare the positions of our candidate ``extra-tidal'' stars with the orbital tracks of the galaxies, and identify stars, both with and without radial velocity matches, that are consistent with lying along the orbital track of the satellites. However, given the small number of candidate stars, we cannot make any conclusive statements about the significance of these spatially correlated stars. Cross matching with publicly available catalogs of RR Lyrae, we find one RR Lyrae candidate with $> 50\%$ astrometric probability outside the limiting radius in each of Sculptor and Fornax, two such candidates in Draco, nine in Ursa Minor, seven in Sextans, and zero in Carina. Follow-up spectra on all of our candidates, including possible metallicity information, will help confirm association with their respective dSphs, and could represent evidence for extended stellar halos or tidal debris around these classical dSphs. 
\end{abstract}


\section{Introduction}

\par The structure and kinematics of classical dwarf spheroidal galaxies (dSphs) (Draco, Sculptor, Ursa Minor, Carina, Sextans, Fornax, Leo I and Leo II) have long been a subject of interest~\citep{Mateo:1998,McConnachie12,2013NewAR..57...52B}. Radial velocity dispersion measurements~\citep{2007ApJ...667L..53W} show that dSphs are amongst the most dark matter-dominated galaxies known, and they may be used as a probe of the nature of dark matter on the smallest scales~\citep{2017ARA&A..55..343B}. Though astrometric data is not yet as sensitive to internal dispersions as radial velocity data~\citep{Strigari:2018bcn}, these data sets have potential to provide significantly more information on the dSphs dark matter halos than radial velocities alone. Previous studies have largely focused on the kinematics and structure of dSphs by studying the stars within their main bodies, i.e. their stellar limiting radii. 

\par Models of the kinematics of dSphs typically assume that these systems are in dynamical equilibrium. However, since they have been accreted into and evolved within the Milly Way (MW) potential, it is certain that they have loss some amount of mass due to tidal stripping. For subhalos that host luminous galaxies, tidal stripping may be imprinted in the observed phase space distribution of the stars. This stripping is most prominent in the Sagittarius dwarf galaxy, which has tidal tails that extend hundreds of degrees from the main body \citep{Newberg02, Majewski03, Koposov12, Ibata20}. At much lower luminosities, $\sim$10$^3$ L$_\odot$, the Tucana III ultra-faint dwarf galaxy has associated streams that are likely due to tidal disruption \citep{Drlica-Wagner15, Shipp18, Mutlu-Pakdil18, Li18}. There are possible hints of tidal disruption of other satellite galaxies, but the evidence remains inconclusive \citep{Carlin18, Mutlu-Pakdil19, Mutlu-Pakdil20,2020MNRAS.496.1092G,2021arXiv210710849L}.

\par Though they do not yet exhibit clear evidence of tidal disruption, photometric and radial velocity studies have been effective in identifying stars associated with Carina, Ursa Minor, and Draco at radii beyond their classical King limiting radius~\citep{2005ApJ...631L.137M,2006ApJ...649..201M}. For the classical dSphs that have been modeled, there is a wide degree of mass loss scenarios possible that still lead to a galaxy that is visible in its current state~\citep{Munoz:2007jn,Penarrubia:2008mu,2015MNRAS.454.2401B,2015NatCo...6.7599U,Wang17}. Stars may be stripped off of these galaxies, though at this time the surface brightness of the associated streams could be well below photometric limits of their detection \citep{Wang17,Genina20}.

\par In recent years, deep photometric measurements of dSphs~\citep{Munoz18,2019ApJ...881..118W} and kinematic data from the {\it Gaia} satellite~\citep{Gaia18} have substanitally improved our understanding of the present state and evolution of the dSphs. 
From the {\it Gaia} data, the proper motions of the entire population of dSphs have become available, leading to an improved characterization of their full three-dimensional motions~\citep{McConnachie_Venn20,Simon18,Fritz18,Pace&Li19,Gaia18}. In addition to the bulk motions, {\it Gaia} EDR3 data have led to, in some cases, the first identification of a small rotation, or streaming, component of the stars in the classical dSphs~\citep{Martinez-Garcia21}. Though the bulk motions and the streaming components may be extracted, the internal velocity dispersions of $\sim$10~km/s are still unable to be resolved by~\gaia~data in itself (see however \citealt{Massari18,Massari20}). 

\par Though the bulk motions of the dSphs are well-characterized via the photometry and kinematics of stars within their main bodies, less well understood are their stellar associations at large radii near the stellar tidal limit of these systems. Identifying member stars at large radii is important, as it can shed light on the extent of the dark matter halo of the dSphs, or possibly determine whether the systems are undergoing tidal disruption \citep{Deason21}. In addition, it may be the first indication of an extended stellar halo associated with dSphs, which would yield important information on their star formation histories. 

\par With this motivation, in this paper we use~\gaia~EDR3 to study the kinematics of a population of six classical dSphs: Draco, Sculptor, Ursa Minor, Carina, Sextans, Fornax. These dSphs are chosen as a representative sample because of their relative proximities and the high-quality astrometry for their brightest stars. In addition, the deduced early infall times~\citep{Rocha:2011aa,Fillingham19} make them good candidates in which these effects may be observed. Since our method for identifying stars at large radii motivated by the Gaia astrometry, it provides an independent and powerful new probe of member stars at large radii as compared to the photometric studies discussed above.

\par This paper is organized as follows. In Section~\ref{sec:data} we review the~\gaia~data and our selection criteria. In Section~\ref{sec:methods} we discuss our methodology for identifying member stars in dSphs. In Section~\ref{sec:results} we present our results for the number of stars associated with each dSph at large radius, cross matches with radial velocity and RR Lyrae surveys, and a discussion of the position of our candidates with respect to the orbital motion of each dSph. In Section~\ref{sec:conclusions} we present our discussion and conclusions.

\section{Data Selection}
\label{sec:data}

\par Expanding upon earlier data releases from the~\gaia~mission \citep{Gaia16},~\gaia~Early Data Release 3 (EDR3, \citealt{Gaia20}) has substantially reduced the statistical uncertainties for individual stellar proper motions (PMs) and reduced systematic uncertainties for the full PM catalog. Here we use the EDR3 data to study regions of interest around six classical dSphs, with a goal of identifying candidate member stars at the largest possible projected radii from them.

\par The  properties of the six classical dSphs that we study are shown in Table~\ref{tab:properties_lit}. We choose these six dSphs because they are the optimal combination of the nearest and the brightest classical dSphs. As indicated, all systems have very similar ellipticities, and their half-light radii vary in a range of $\sim 0.1 - 1$ kpc. The radial velocities (RVs) and the velocity dispersions for these dSphs are well measured, and estimates of their mass distributions from these data sets indicate that they all are dark matter dominated~\citep{Battaglia:2013}. Only for Draco and Sculptor~\citep{Massari18,Massari20} are internal tangential velocity dispersions measured, though they are substantially less well-constrained than the RV measurements.

\par To select a candidate sample of stars associated with each dSph, we perform a series of cuts through a sequence of several steps. First, we apply a cut to select sources in EDR3 which are within a fixed projected radius from the center of each dSph. Specifically, we select stars within a projected radius of 3$^{\circ}$ around each dSph. This radius corresponds to around 10 times the projected half-light radius for all of the dSphs, which provides sufficient sample size to characterize the populations of the main body and the MW foreground. 

Once these cuts have been implemented, to produce an astrometrically well-behaved sample we consider only stars with reliable astrometric solutions. Specifically, we keep only sources that pass the following EDR3 cut criteria:
\begin{itemize}
\item $|C^*| \leq 3\sigma_{C^*} (G)$
\item $\texttt{ruwe} < 1.4$ 
\item $\texttt{ipd\_gof\_harmonic\_amplitude} \leq 0.2$ 
\item $\texttt{visibility\_periods\_used} \geq 9 $ 
\item $\texttt{astrometric\_excess\_noise\_sig} \leq 2$ 
\item $\texttt{ipd\_frac\_multi\_peak} \leq 2$ 
\item $\texttt{astrometric\_params\_solved} > 3$
\item no~\texttt{duplicated\_source}~in EDR3.
\end{itemize}
Here $|C^*|$ is the corrected BP and RP flux excess factor (as defined in Equation 6 of \cite{Riello21}), $\sigma_{C^*} (G)$ is defined in Equation 18 of \cite{Riello21}, and $\texttt{ruwe}$ is the renormalised unit weight error provided in EDR3 catalogue. We then remove any known active galactic nucleus (AGN) in \gaia\ EDR3 \citep{Gaia20} to reduce contamination. We correct the \emph{G}-band magnitude and the flux excess factor as suggested in~\cite{Riello21}, and the Gaia dust extinctions for $G$ and $G_{RP}$ assuming relations from  \cite{Babusiaux18}.

\par In addition to the cuts above, we perform additional cuts that are more specific to the nature of our analysis. First, we remove stars that are obvious members of the foreground by removing stars with well-measured non-zero parallaxes. This is implemented with a parallax cut as $\omega - 3\sigma_{\omega} < 0$, where $\omega$ is the parallax and $\sigma_w$ is the error on the parallax. On top of this parallax cut, we perform a second cut by removing stars that would have a transverse velocity larger than the escape velocity of the MW at the distance of each dSph. We used the \texttt{MWPotential2014} potential in \texttt{galpy} \citep{galpy15} to determine the escape velocity ($V_{esc}$) at a dSph's given distance. Specifically, we only keep stars with $4.74 D_\odot (PM - 3\sigma_{PM})  \leq V_{esc}$, where PM is the proper motion in the plane of the sky, and $\sigma_{PM}$ is the error on the PM.

\par
\begin{figure*}
\includegraphics[width=1.0\textwidth]{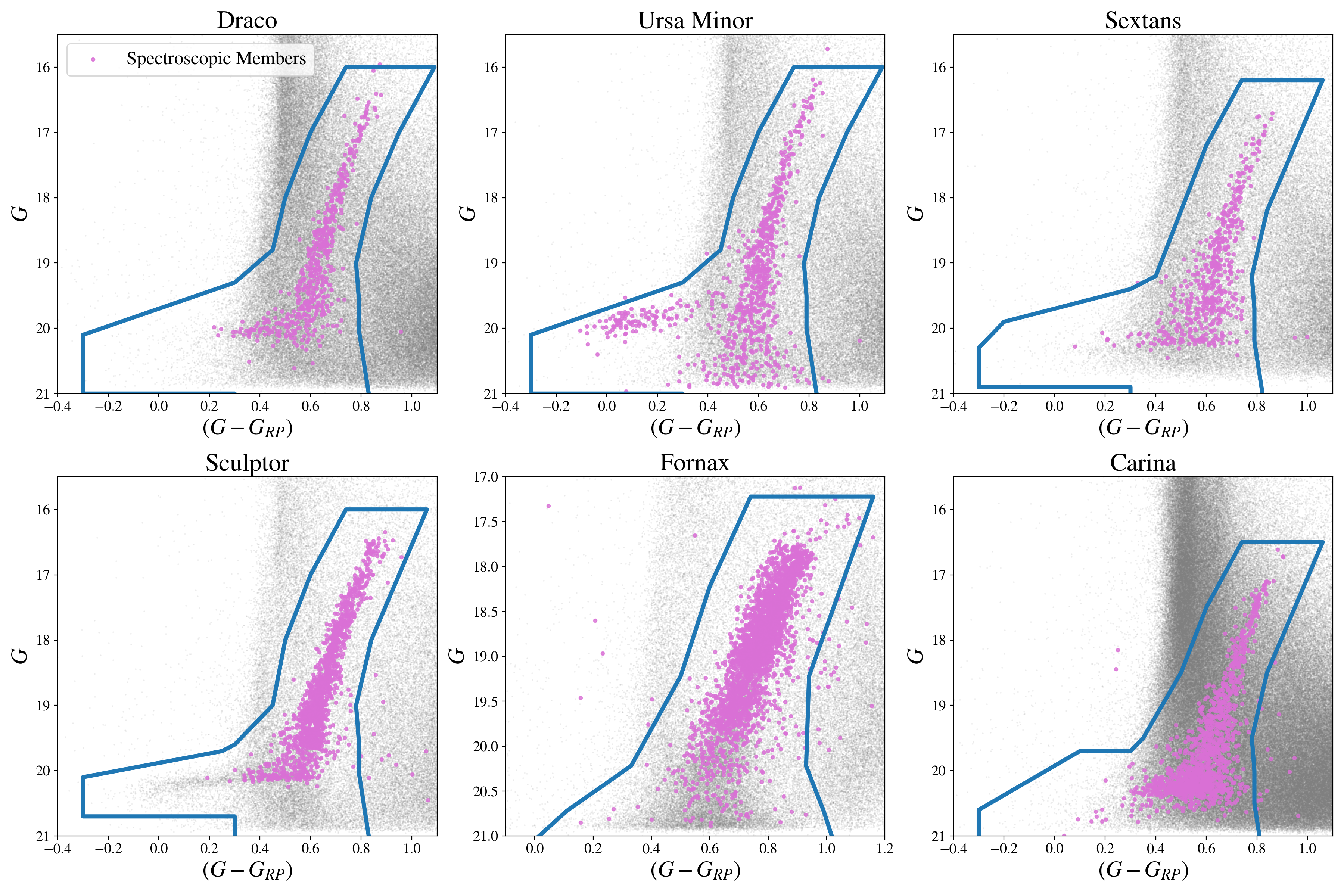}
\caption{Color-magnitude diagrams of the six satellite galaxies examined in this study. Grey points indicate all \gaia\ EDR3 stars within 3 degrees of each satellite. Purple points indicate stars from publicly available radial velocity (RV) catalogs that have been cross-matched with EDR3 and are consistent with the satellite for its systemic RV. Finally, the boundary used for trimming our EDR3 sample to remove contaminating Milky Way stars is marked by the dark blue line. 
}
\label{fig:data_cmd}
\end{figure*}

 After the parallax cut, our final step to minimize MW contamination in our sample is to select stars based on their position in color-magnitude space.
To create this cut, we start by compiling publicly available spectroscopic data for each dSph \citep{Walker09, Walker15, Battaglia11, Hendricks14,  Fabrizio16,Spencer18,Pace20,Pace21} and cross-match these catalogs with EDR3.
Then, for each dSph spectroscopic catalog, we keep stars whose RVs are consistent within three sigma of the systemic RV ($V_r$) of the dSph (where sigma is the velocity dispersion of the system, as listed in Table \ref{tab:properties_lit} along with $V_r$).
We plot this trimmed sample in \gaia\ color-magnitude space using $G$ and $G-G_{RP}$ (as noted in \citet{Gaia20}, the $G_{BP}$ filter suffers from bias at faint magnitudes so we choose to use only $G$ and $G_{RP}$ given many of our targets are near the \gaia\ faint limit).
Finally, we use the position of the spectroscopically-selected stars to define an area in the color-magnitude diagram (CMD) that is consistent with the dSph (as seen in Figure \ref{fig:data_cmd} for all six satellites).
\par After implementing all of these cuts, our final samples of stars from~\gaia~for each dSph are as follows: Draco (10799), Ursa Minor (6215), Sextans (6219), Sculptor (8340), Fornax (21072), and Carina (23498).




\tabcolsep=0.18cm
\begin{table*}
\centering
\caption{Properties of each of the dSphs in our sample. The columns are: (1) galaxy name; (2) and (3): position \citep{McConnachie_Venn20b}; (4) magnitude; (5) distance; (6) position angle based on Plummer models; (7) ellipticity; (8) Plummer half-light radius (in arcmin); (9) Plummer half-light radius (in kpc for the reader's reference); (10) King limiting radius, with the exception of Fornax using \citet{Battaglia06}; (11) King limiting radius (in kpc for the reader's reference);  (12) radial velocity; (13) radial velocity uncertainties. (4) to (11) are from \citet{Munoz18}. (12) and (13) are from \citet{Simon19}.
}

\begin{tabular}{lrrrrrrrrrrrr} 
\hline
Galaxy & $\alpha$ & $\delta$ & $M_V$  & $D_\odot$ & $\Theta_p$ & $\epsilon$ & $r_{h,p}$  & $r_{h,p}$  & $r_{t,k}$ & $r_{t,k}$  & $V_r$ & $\sigma_{v}$\\
 & [deg] & [deg] &   & [kpc] & [deg] &  & [arcmin] & [kpc] & [arcmin] & [kpc] & [${\rm km~s^{-1}}$] & [${\rm km~s^{-1}}$] \\
\hline
Draco & $260.0517$ & $57.9153$ & $-8.71\pm0.05$ & $76$ & $87$ & $0.29$ &  $9.67\pm0.09$ &$0.214$ &   $48.1\pm1.3$ & $1.064$ & $-291.7$  & $9.1$\\
Ursa Minor & $227.2854$ & $67.2225$ & $-9.03\pm0.05$ & $76$ & $50$ & $0.55$ & $18.3\pm0.11$ &$0.407$ & $77.3\pm0.7$ & $1.709$ & $-247.2$ & $9.5$\\
Sextans & $153.2625$ & $-1.6147$ & $-8.72 \pm 0.06$ & $ 86$ & $57$ & $0.30$ & $16.5\pm0.10$ &$0.413$ & $60.5\pm0.6$ & $1.514$ & $224.3$ & $7.9$\\
Sculptor & $15.0392$ & $-33.7092$ & $-10.82 \pm 0.14$ & $ 86$ & $92$ & $0.33$ & $11.17\pm0.05$ &$0.280$ & $74.1\pm0.4$ & $1.854$ & $111.4$ & $9.2$\\
Fornax & $39.9971$ & $-34.4492$ & $-13.46\pm0.14$ & $147$ & $45$ & $0.29$ & $19.6\pm0.08$ &$0.838$ & $69.1\pm0.4$ & $2.955$ & $55.2$ & $11.7$\\
Carina & $100.4029$ & $-50.9661$ & $-9.43\pm0.05$ & $105$ & $60$ & $0.36$ & $10.1\pm0.10$ &$0.308$ & $58.4\pm0.98$ & $1.784$ & $222.9$ & $6.6$\\
\hline
\end{tabular}
\label{tab:properties_lit}
\end{table*}

\section{Likelihood analysis}
\label{sec:methods}
We now discuss our methodology for assigning membership probabilities to stars in each of our dSphs using both astrometric and photometric data. We note that~\gaia~radial velocities are not available for the stars in our sample, as they are all fainter than the magnitude limit of the DR2 radial velocity sample. The magnitude limit of Gaia RVs is $G \sim 14$ which is brighter than the tip of the red-giant branch (TRGB) in every galaxy, and Gaia DR2 RVs are brighter than this. As we discuss below, while we achieve accurate membership results using PMs alone, we include information on the spatial distribution through photometry along with the PM data to improve upon the model and further eliminate the influence of MW foreground stars. The methods discussed here follow those established in~\citet{Pace&Li19}, we review the salient aspects for our analysis. 

\par For stars in our catalogue, we adopt the orthographic projection of RA and DEC and PMs from~\cite{Gaia18}, and transform them into the Cartesian frame as
\begin{equation}\label{eq_xy_Coord} 
\begin{split}
x & = \cos\delta \sin(\alpha -\alpha_C)\\
y & = \sin\delta \cos\delta_C -\cos\delta \sin\delta_C \cos(\alpha - \alpha_C)\\
\end{split}
\end{equation}
where $\alpha$ and $\delta$ are the location of a star in the RA and DEC direction, and ($\alpha_{C}$, $\delta_{C}$) is the center of the dSph. After completing the transformation, we begin our analysis by assuming our catalogue is composed of two populations: a general MW foreground/background population and members of the dSph. As in~\cite{Pace&Li19}, we define the total likelihood function as:
\begin{equation}
    \mathcal{L} = f_{sat} \mathcal{L}_{sat} + (1-f_{sat}) \mathcal{L}_{MW}. 
\end{equation}
The likelihood is composed of two components. First, the component $\mathcal{L}_{sat}$ describing the stars that belong to the dSph, and second, the component $\mathcal{L}_{MW}$ describing the MW foreground/background stars. The quantity $f_{sat}$ is defined as the fraction of stars that belong to the dSph, so that $1-f_{sat}$ represents the fraction of stars that belong to the MW foreground/background. 

\par The likelihood components $\mathcal{L}_{sat}$ and $\mathcal{L}_{MW}$ can be broken up into two components, one that depends on the kinematics through the PMs, and one that depends on the spatial distribution. Adding in these components, the likelihood takes the form, 
\begin{equation}
\label{eq_tot_likelihood}
\begin{split}
    \mathcal{L} & = f_{sat} \mathcal{L}_{sat,spatial} \mathcal{L}_{sat,PM} \\
    & + (1-f_{sat}) \mathcal{L}_{MW,spatial} \mathcal{L}_{MW,PM} \\
\end{split}
\end{equation}
where $\mathcal{L}_{sat,spatial}$ and $\mathcal{L}_{sat,PM}$ corresponds to the photometric and proper motion likelihoods for the dSph, and $\mathcal{L}_{MW,spatial}$ and $\mathcal{L}_{MW,PM}$ are the respective likelihood functions for the MW component. 

\par For each of the dSph and MW PM components, we assume both follow independent multi-variate gaussian distributions \citep{Vasiliev19b}:  
\begin{equation} 
\begin{split}
\mathcal{L}_{\imath,PM} & = (2\pi)^{-N/2}(\textrm{det} \Sigma)^{-1/2} \\
& \times\exp[-\frac{1}{2}(\mu_\imath - \mu_\imath^{mod})^T\Sigma^{-1}(\mu_\imath - \mu_\imath^{mod})] \\
\end{split}
\end{equation}
where $N$ is the total number of the stars in the sample, which are given for each dSph in Section \ref{sec:data}, and $\imath = $ (sat, MW) refers to either the dSph or the MW foreground/background component. Here $\mu_\imath$ represents the EDR3 measured PM vectors of stars that belong to a given component, and $\mu_\imath^{mod}$ gives the expected value of the PM, and the co-variance matrix $\Sigma$ is defined as:
\begin{gather}
\Sigma = 
\begin{bmatrix} E_1+S_1 & 0_{2\times2} & \ldots & 0_{2\times2} \\ 0_{2\times2} & E_2+S_2 & \ldots \\ \vdots & \vdots & \ddots & \vdots \\ 0_{2\times2} & 0_{2\times2} & \ldots & E_N+S_N \end{bmatrix}
\end{gather}
with
\begin{gather}
E_i = 
\begin{bmatrix}
\epsilon_{i,x}^2 & \rho_i\epsilon_{i,x}\epsilon_{i,y} \\
\rho_i\epsilon_{i,x}\epsilon_{i,y} & \epsilon_{i,y}^2
\end{bmatrix}
\, \, , \, \, 
S_i = 
\begin{bmatrix}
\sigma_i^2(x_i) & 0 \\ 0 & \sigma_i^2(x_i) 
\end{bmatrix}
\end{gather}
where $\rho_i$ is the EDR3-defined correlation between $\mu_{\alpha\star} = \mu_{\alpha} \cos \delta$ and $\mu_{\delta}$ for each star, $\epsilon_{i,x}$ and $\epsilon_{i,y }$ are uncertainties on $\mu_{\alpha\star}$, $\mu_{\delta}$ correspondingly, and $\sigma_i$ is the modeled velocity dispersion at the location of each star.  
 
\par We note that the assumption of a multi-variate gaussian is likely an oversimplification for both populations, in particularly for the MW population. As we discuss below, we test the general properties of this model by comparing to the photometric data using only the kinematic components of the likelihood, and from this analysis we find good agreement with existing photometric data. For this reason we have confidence that this approach is reasonable given the current sensitivity of~\gaia. 

\par For the photometric model, we assume that the projected density of each dSph is described by a  Plummer profile~\citep{Plummer1911}
\begin{equation}
\Sigma(R_e) = \frac{1}{\pi a^2_h (1 - \epsilon)} (1 + R^2_e / a^2_h)^{-2}
\end{equation}
where $a_h$ is the semi-major half-light radius, $\epsilon$ is the projected ellipticity, and $R_e$ is the elliptical radius defined as
\begin{equation}
R^2_e = R_x^2 + \frac{ R_y^2}{(1 - \epsilon)^2}.
\end{equation}
Here we rotate $R_x$ and $R_y$ by the position angle $\Theta$, which is measured North to East in the frame of the sky:
\begin{equation}
\label{eq_XY_t}
\begin{split}
R_x & = x \cos(\Theta) - y \sin(\Theta)\\
R_y & = x \sin(\Theta) + y \cos(\Theta)\\
\end{split}
\end{equation}
where $x$ and $y$ are defined in Equation \ref{eq_xy_Coord}. 
\par From the Plummer model, the normalized version of the spatial component of the likelihood function is given by 

\begin{equation}
\begin{split}
\mathcal{L}_{sat,spatial} & = \frac{\sqrt{\displaystyle(a^2_h + R^2_{\mathrm{max}})((1 - \epsilon)^2 a^2_h + R^2_{\mathrm{max}})}}{ 1 - \epsilon} \\
& \times \frac{\displaystyle a^2_h}{\displaystyle R^2_{\mathrm{max}} (a^2_h + R^2_e)^2}\\
\end{split}
\end{equation}

With the assumption of a constant spatial distribution for the MW population, $\mathcal{L}_{MW,spatial} = \frac{1}{R^2_{\mathrm{max}}}$
~, the spatial likelihood acts as a weighting factor preferentially selecting stars that are closer to the center of the dSph. Note that in our analysis we have $R_{max} = 3^\circ$. 

\par We use the likelihood defined above to estimate model parameters. To estimate the best value for the parameters, we use the~\texttt{emcee}~library \citep{Foreman-Mackey13}, which implements the affine-invariant ensemble sampler for Markov Chain Monte Carlo (MCMC; \citealt{Goodman&Weare10}) in Python. We free up to 10 parameters and their corresponding priors in the analysis, which are: 
\begin{itemize}
    \item One uniform prior between 0 and 1 for the fraction of the stars belonging to each dSph: $f_{sat}$
    \item Three parameters describing the spatial distribution: one log prior 
    for ${\rm log}_{10}$($\frac{R_e}{1~deg}$) between -2 and 0; one uniform prior between 0 and 1 for ellipticity; one uniform prior between -90 and 90 degrees for position angle
    \item Two uniform priors between -1 to 1 ${\rm mas~yr^{-1}}$ for parameters describing the dSph PM: ($\mu_{\alpha, \star} \equiv \mu_{\alpha} cos\delta$, $\mu_{\delta}$)
    \item Two uniform priors between -5 and 5 ${\rm mas~yr^{-1}}$ for parameters describing the MW systematic PM:  ($\mu_{\alpha, \star}^{MW} \equiv \mu^{MW}_{\alpha}cos\delta$, $\mu^{MW}_{\delta}$)
    \item Two log uniform between -2 and 1 ${\rm mas~yr^{-1}}$ for parameters describing the MW proper motion dispersion: ($\sigma^{MW}_{\mu_{\alpha} cos\delta}$ and $\sigma^{MW}_{\mu_{\delta}}$)
\end{itemize}

\par We fix the parameters that represent the intrinsic proper motion velocity dispersions of the dSphs to the values that are determined from their measured radial velocity dispersions in Table~\ref{tab:properties_lit}. Note that this assumes that the velocity dispersions of all the dSphs are isotropic, but given that~\gaia~data in itself is not sensitive to the internal PM velocity dispersions, this assumption does not impact our analysis.
For the position angle, the prior range is set based on the conventions in our Cartesian projection; however for final reporting, we present our best-fit position angle in the standard literature convention of defining the position angle from North over East.

\par From the posterior of the~\texttt{emcee}~runs, we calculate the probability,  $P_{i,\text{PM+Spatial}}$, that the $i^{th}$ star belongs to the dSph as
\begin{equation}
\label{eq_member_prob_spatial}
P_{i,\text{PM+Spatial}} =  \frac{f_{sat;i}  \mathcal{L}_{sat,spatial;i} \mathcal{L}_{sat,PM;i}}{\mathcal{L}_{i}}
\end{equation}
where $\mathcal{L}_{i}$ is the $i^{th}$ value in Equation \ref{eq_tot_likelihood}.
This provides us a probability distribution of $P_{i,\text{PM+Spatial}}$ of each star, which we then use to calculate the membership possibility by finding median of its $P_{i,\text{PM+Spatial}}$ distribution. Following the same method, we can compute the PM only probability, $P_{i,PM}$, for individual star $i$ as:
\begin{equation}
\label{eq_member_prob}
P_{i,PM} =  \frac{f_{sat;i} \mathcal{L}_{sat,PM;i}}
{f_{sat;i}  \mathcal{L}_{sat,PM;i} + (1-f_{sat;i})\mathcal{L}_{MW,PM;i}}
\end{equation}

\begin{table*}
\centering
\caption{Properties of each of the dSphs generated by MCMC. The columns are: (1) galaxy name; (2) and (3): proper motion; (4) half-light radius; (5) ellipticity; (6) position angle (measured North to East).
}

\label{tab:Properties}
\begin{tabular}{lrrrrrrr} 
\hline
Galaxy & $\mu_{\alpha}cos\delta$ & $\mu_{\delta}$ & $a_H$ & $\epsilon$ & $pa$\\
& [${\rm mas~yr^{-1}}$] & [${\rm mas~yr^{-1}}$] & [arcmin] &  & [deg] \\
\hline
Draco  & $0.045\pm0.006$ & $-0.188\pm0.006$ & $8.97\pm0.25$ & $0.29\pm0.02$ & $90.73\pm2.73$\\
Ursa Minor  & $-0.121\pm0.005$ & $0.073\pm0.005$ & $18.83\pm0.48$ & $0.49\pm0.02$ & $51.16\pm1.18$\\
Sextans & $-0.404\pm0.009$ & $0.034\pm0.009$ & $22.87\pm0.77$ & $0.25\pm0.03$ & $53.35\pm3.57$\\
Sculptor & $0.100\pm0.003$ & $-0.157\pm0.002$ & $12.05\pm0.16$ & $0.30\pm0.01$ & $94.66\pm1.16$\\
Fornax & $0.378\pm0.001$ & $-0.352\pm0.002$ & $18.50\pm0.14$ & $0.33\pm0.01$ & $39.75\pm0.59$\\
Carina & $0.538\pm0.006$ & $0.125\pm0.006$ & $10.64\pm0.28$ & $0.37\pm0.02$ & $63.17\pm1.94$\\
\hline
\end{tabular}
\end{table*}

\section{Results}
\label{sec:results} 
We now move to on presenting our results for the membership probabilities for each dSph. We begin by focusing on the results obtained from the likelihood method described above. We then move on to cross-match our highest probability stars with radial velocity measurements from each dSph and from RR Lyrae catalogs. For stars that we identify as highly-probable members at large radii, we then compare to the orbital trajectories of each dSph to determine how the stars at large radii compare to the best-fitting orbits. 

\par Before considering membership probabilities, to calibrate to previous results we first measure the systemic PM for each dSph. Specifically, we compare to the work of~\citet{Gaia18} and~\citet{McConnachie_Venn20} for DR2, and the updated EDR3 systemic PMs from~\citet{McConnachie_Venn20b} ,~\citet{Battaglia21}, and \citet{Martinez-Garcia21}. 
All of these works use a selection radius of $\lesssim 2$ degrees around each dSph, so calibrating to these results provides a test for the 3$^\circ$  cut radius implemented in our analysis. 

\par
\begin{figure*}
\includegraphics[width=1\linewidth]{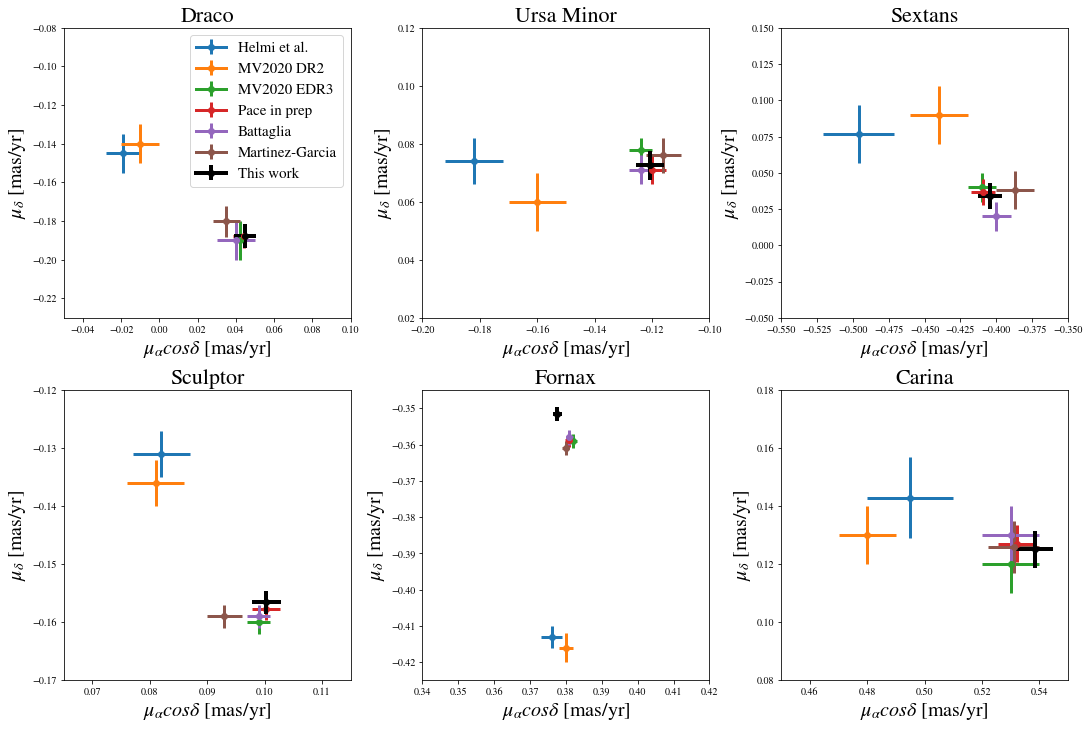} \\
\caption{Comparison of systemic proper motions of Draco, Ursa Minor, Sextans, Sculptor, Sextans, and Carina in $\mu_{\alpha} \cos\delta$ and $\mu_{\delta}$ directions. The black error bars shows PM with $1\sigma$ uncertainties using \gaia EDR3. Blue error bars show \citet{Gaia18}'s value and uncertainty of PMs using \gaia DR2. Orange error bars show the PMs of \citet{McConnachie_Venn20} using \gaia DR2 data, while the green error bars are the updated PMs of \citet{McConnachie_Venn20b} using \gaia EDR3. For other values based on \gaia EDR3, we have error bars in red representing the PMs of Pace et al. in prep. The purple cross show PMs from \citet{Battaglia21}, and the brown cross represent PMs of \citet{Martinez-Garcia21}. We can see that our PM distributions are in good agreement with PM values using \gaia EDR3 data. 
}
\label{fig:pm_comp}
\end{figure*}

 In Figure~\ref{fig:pm_comp}, we show the comparison of PMs. The error bars on our analysis represent the one-sigma containment regions for the posterior of the PM. Note that there is a shift from DR2 to EDR3 in the measured PMs for each dSph, which is mostly due to the reduced systematics between the different data sets. We specifically see that our results are consistent with~\cite{McConnachie_Venn20b}, who also use EDR3 data. The value of the PMs that we obtain are listed in Table~\ref{tab:Properties}.

\subsection{Member identification} 

\par We then move on to examine the membership probabilities. To calculate these, we consider two implementations of our likelihood analysis: one including the photometric and kinematic components ($P_{\text{PM+Spatial}}$, as defined in Equation  \ref{eq_member_prob_spatial}), and one kinematics only component  ($P_{PM}$, as defined in Equation  \ref{eq_member_prob}). The latter implementation is in particular useful to obtain a surface brightness profile for the dSphs that is independent of previous measurements, and as described above this serves as a good test of our multi-variate guassian model. 

\begin{figure*}
\includegraphics[width=1\linewidth]{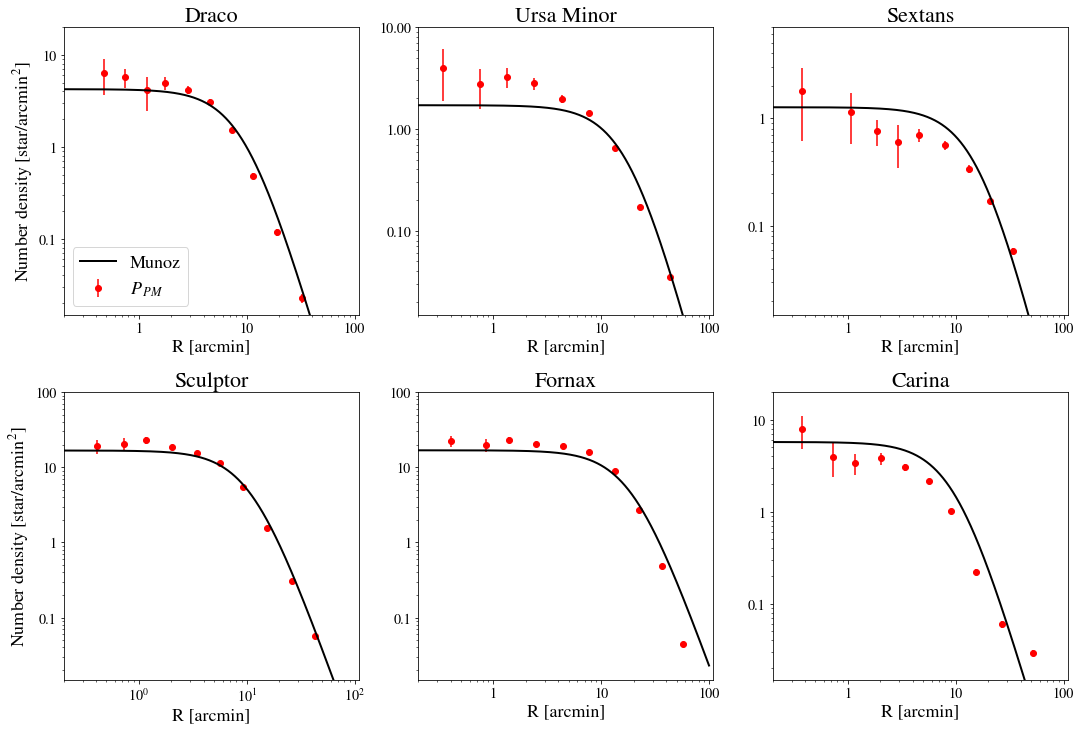} \\
\caption{The surface brightness based on MCMC generated proper motion only membership probability ($P_{PM}$) in red, comparing with the Plummer model in black from \citet{Munoz18}. The number densities in red are calculated using the sum of $P_{PM}$ for every star in the given bin divided by the area, while the error bars represent the square root of the sum of $P_{PM}$ for every star in the given bin divided by the area.}
\label{fig:Luminosity}
\end{figure*}

In Figure \ref{fig:Luminosity}, we plot the surface brightness profiles for dSphs. We sum over $P_{PM}$ for every star within the given radius bin divided by the area to plot the number density in red, along with the Plummer profiles in black using measurements from \cite{Munoz18} (see Table \ref{tab:properties_lit}). The $P_{PM}$ generated surface brightness profiles agree with those obtained in \cite{Munoz18}, which gives us confidence that our multi-variate Gaussian likelihood model is a good description of the systems.

\begin{figure*}
\includegraphics[width=.99\linewidth]{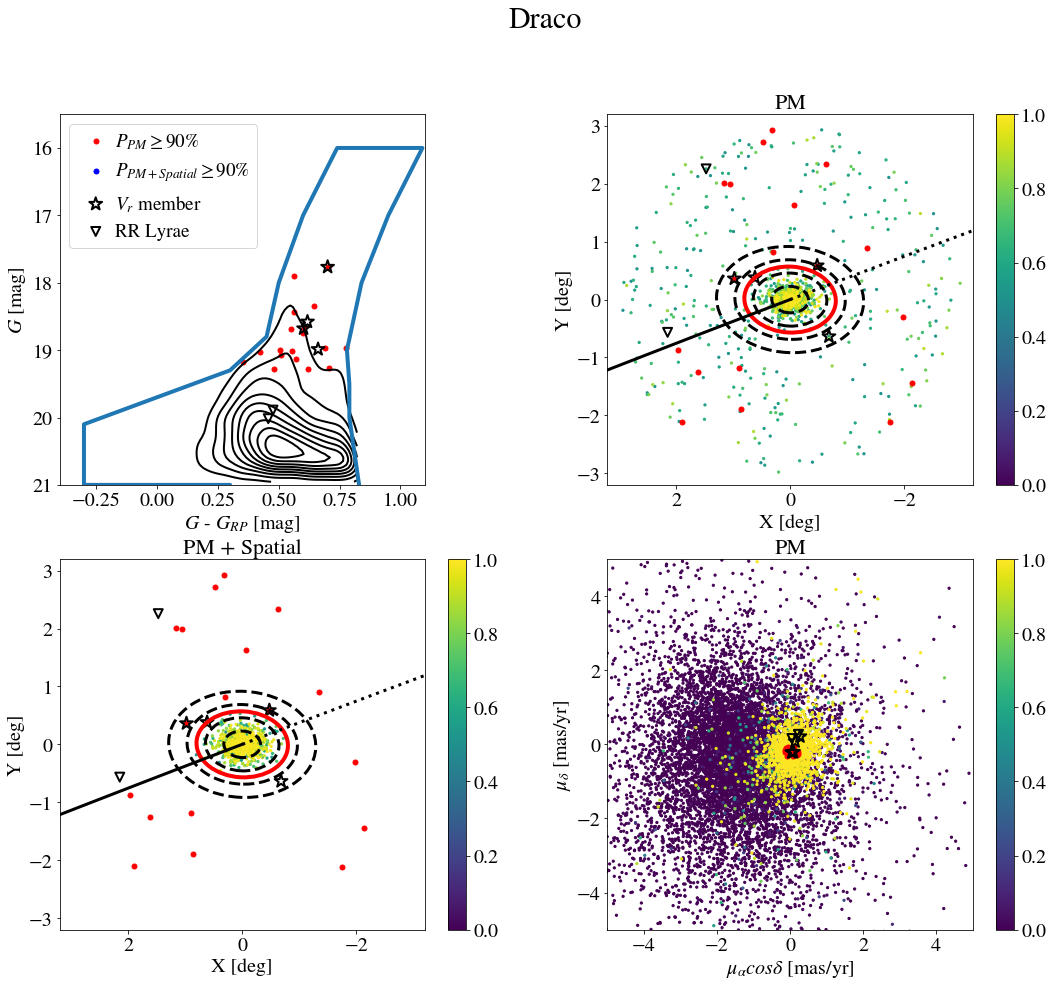} 
\caption{In upper left we have the identified possible extra-tidal candidates ($P_{PM} \geq 90\%$) using proper motion only membership probability (in red) and proper motion plus spatial membership probability ($P_\text{PM+Spatial} \geq 90\%$ in blue) with our CMD selection for Draco. The black contour lines show the density of likely MW foreground/background stars in the CMD selection (stars with $P_{PM} < 1\%$). The star symbols represent the $V_r$ members, which are stars with cross-matched RV that are consistent with the bulk velocity of Draco (RV within $3\sigma_v$ of $V_r$ in Table~ \ref{tab:properties_lit}). The upside down triangles show the cross-matched RR Lyrae that are consistent in distance with Draco. Both $G$ and $G_{RP}$ have been corrected (see Section \ref{sec:data}).
Upper right: Stars with their PM-only membership probability of $P_\text{PM}\geq50\%$ distribution for Draco in X-Y coordinates centered at 0, color-coded with $P_{PM}$ from 0 to 1, labeled with extra-tidal candidates. The red ellipse represents the $r_{t,k}$ transformed by position angle and ellipticity. The black dashed ellipses show $2 r_{h,p}$, $4 r_{h,p}$, $6 r_{h,p}$, $8 r_{h,p}$ transformed by position angle and ellipticity, respectively. We can see that the candidates and higher probability stars are concentrated in the middle. We also show the orbit of the satellite forwards (backwards) in time as a solid (dotted) black line.
Lower left: Stars with their PM + Spatial membership probability of $P_\text{PM+Spatial}\geq50\%$, color-coded with $P_\text{PM+Spatial}$ from 0 to 1, are being shown in X-Y coordinates. 
Lower right: Proper motions distribution for our full \gaia EDR3 catalog for Draco, centered at 0, color-coded with $P_{PM}$ from 0 to 1. We can see that the candidates and higher probability stars are concentrated in the middle.}

\label{fig:Draco}
\end{figure*}

\begin{figure*}
 
  \includegraphics[width=.99\linewidth]{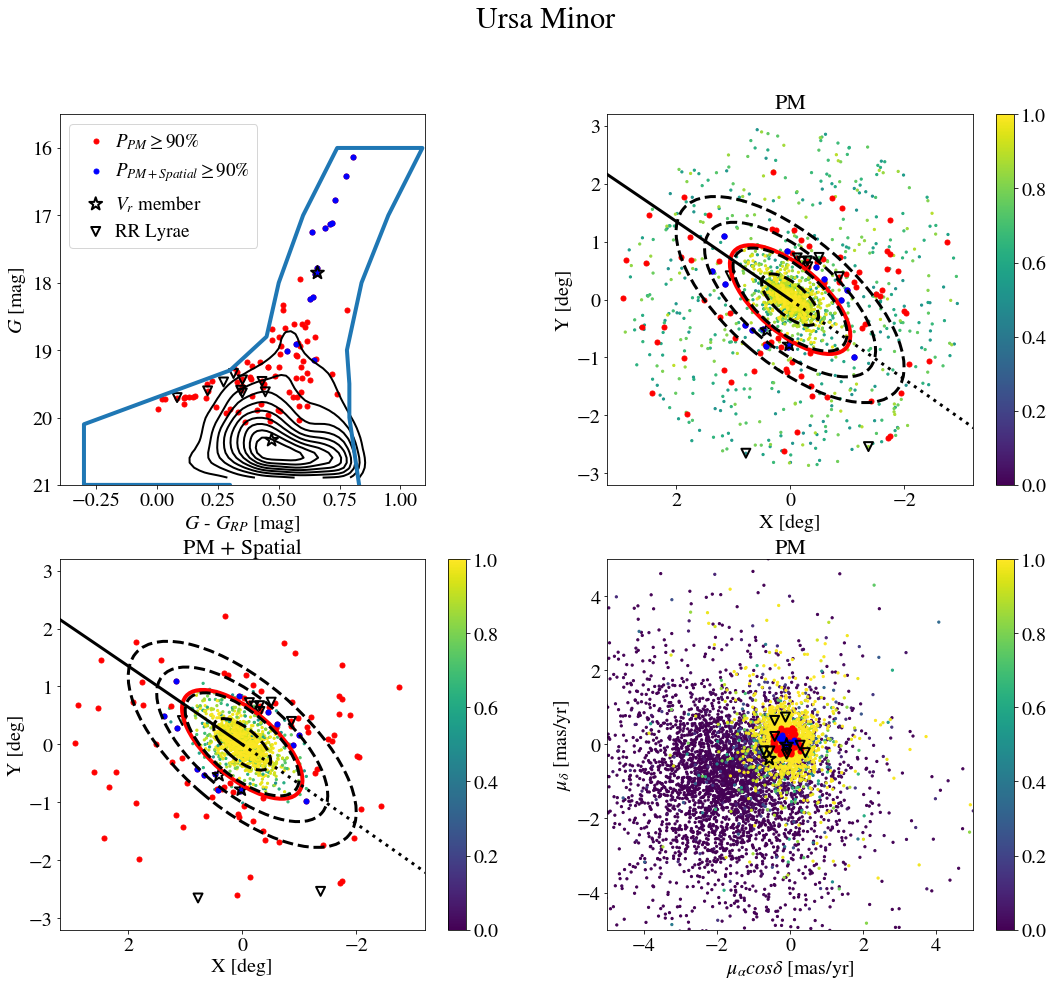} 
  \caption{Same format as Figure \ref{fig:Draco} for Ursa Minor. 
  }
  \label{fig:Ursa_Minor}

 \end{figure*}

\begin{figure*}
  \includegraphics[width=.99\linewidth]{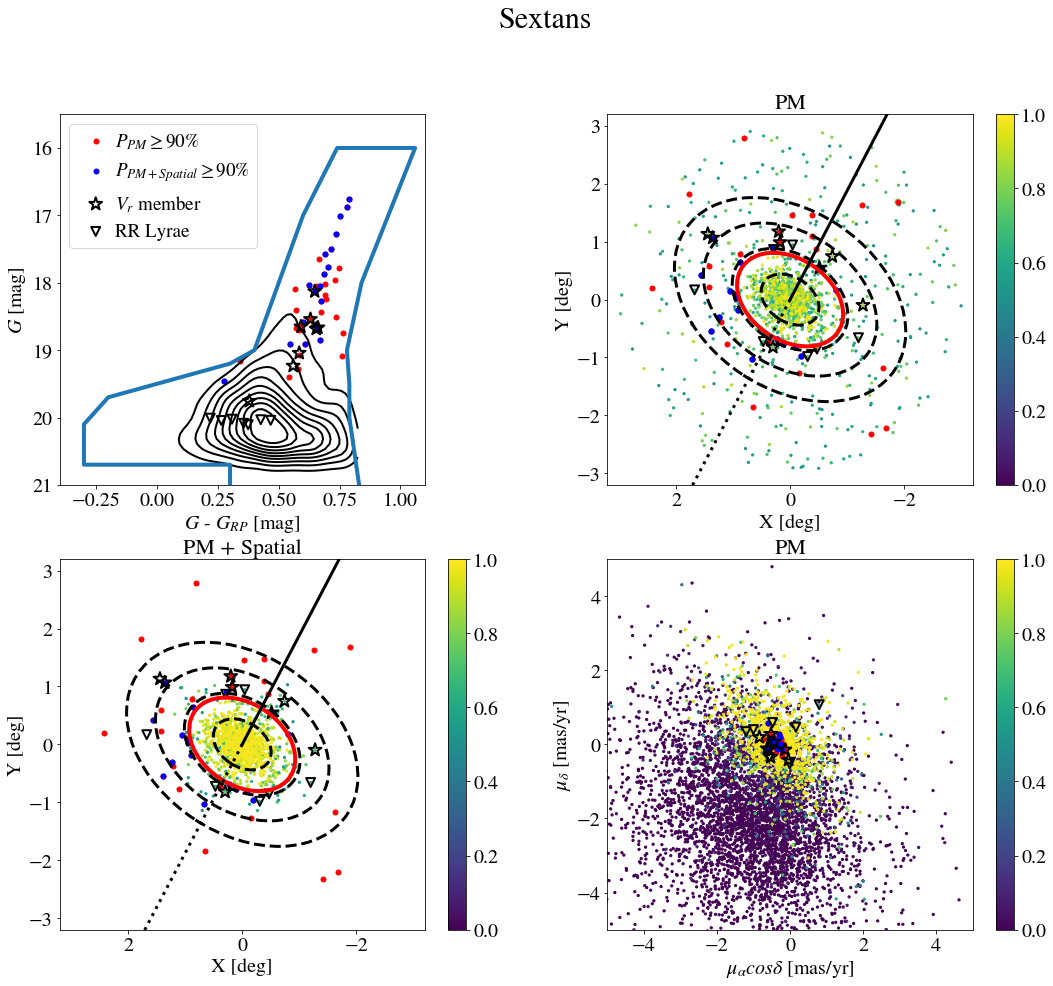} 
  \caption{Same format as Figure \ref{fig:Draco} for Sextans.}
  \label{fig:Sextans} 
 \end{figure*}

\begin{figure*}  \includegraphics[width=.99\linewidth]{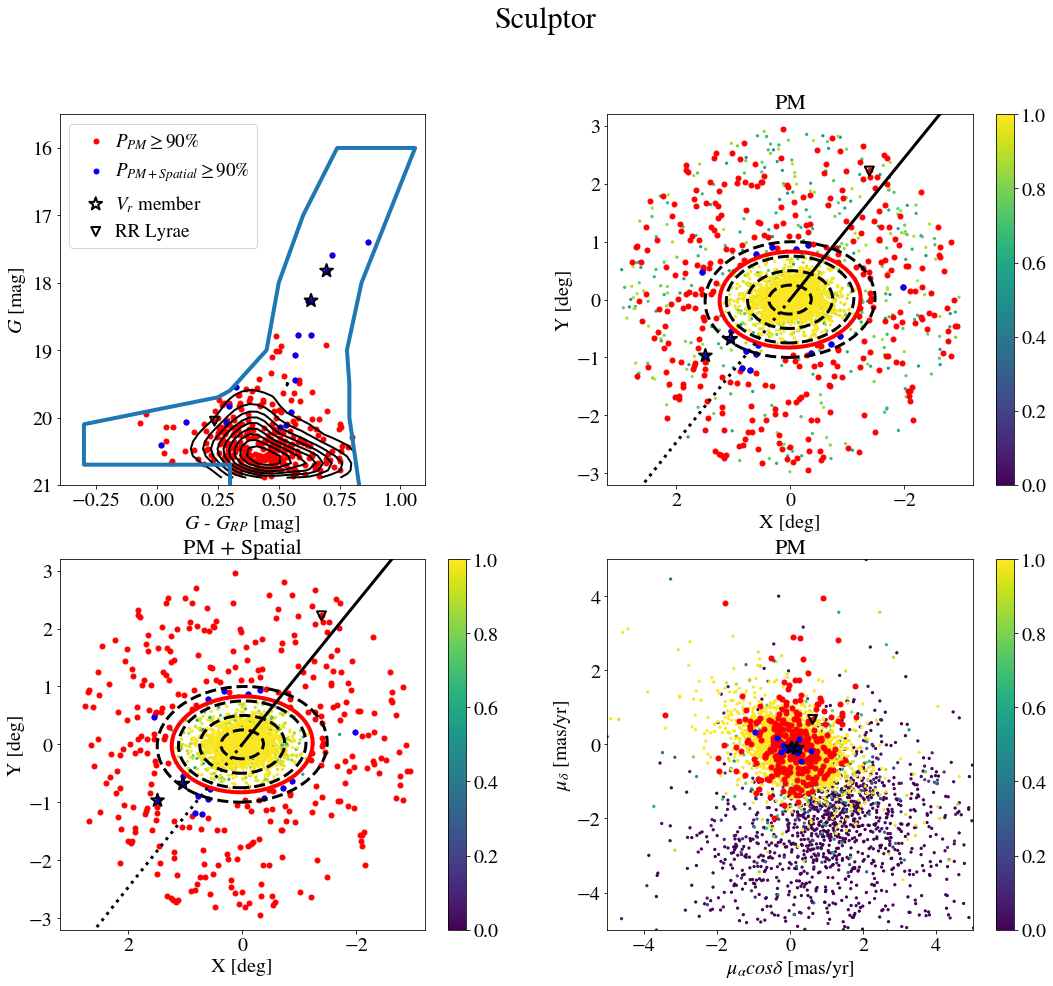} 
  \caption{Same format as Figure \ref{fig:Draco} for Sculptor.}
  \label{fig:Sculptor}

 \end{figure*}

\begin{figure*}
  \includegraphics[width=.99\linewidth]{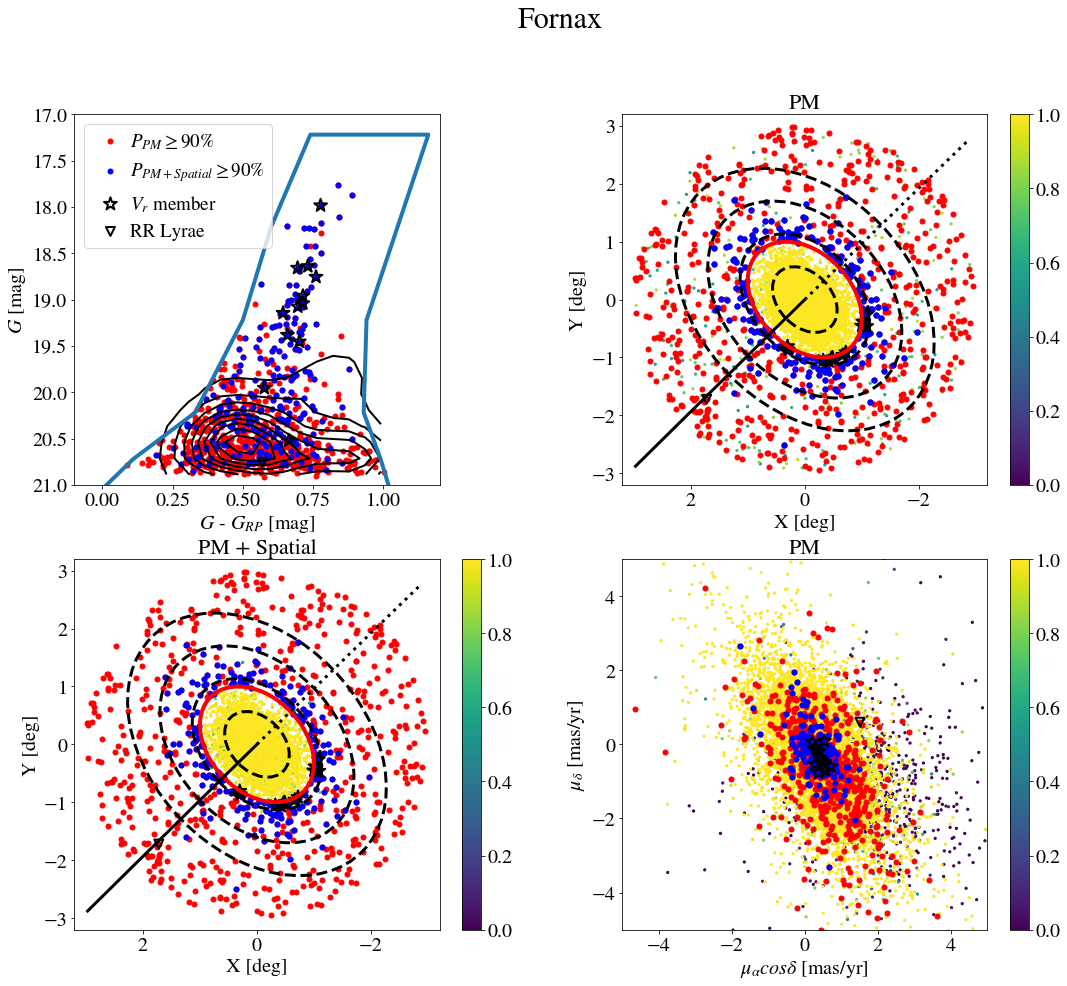} 
  \caption{Same format as Figure \ref{fig:Draco} for Fornax.}
  \label{fig:Fornax}
 \end{figure*}

\begin{figure*}
  \includegraphics[width=.99\linewidth]{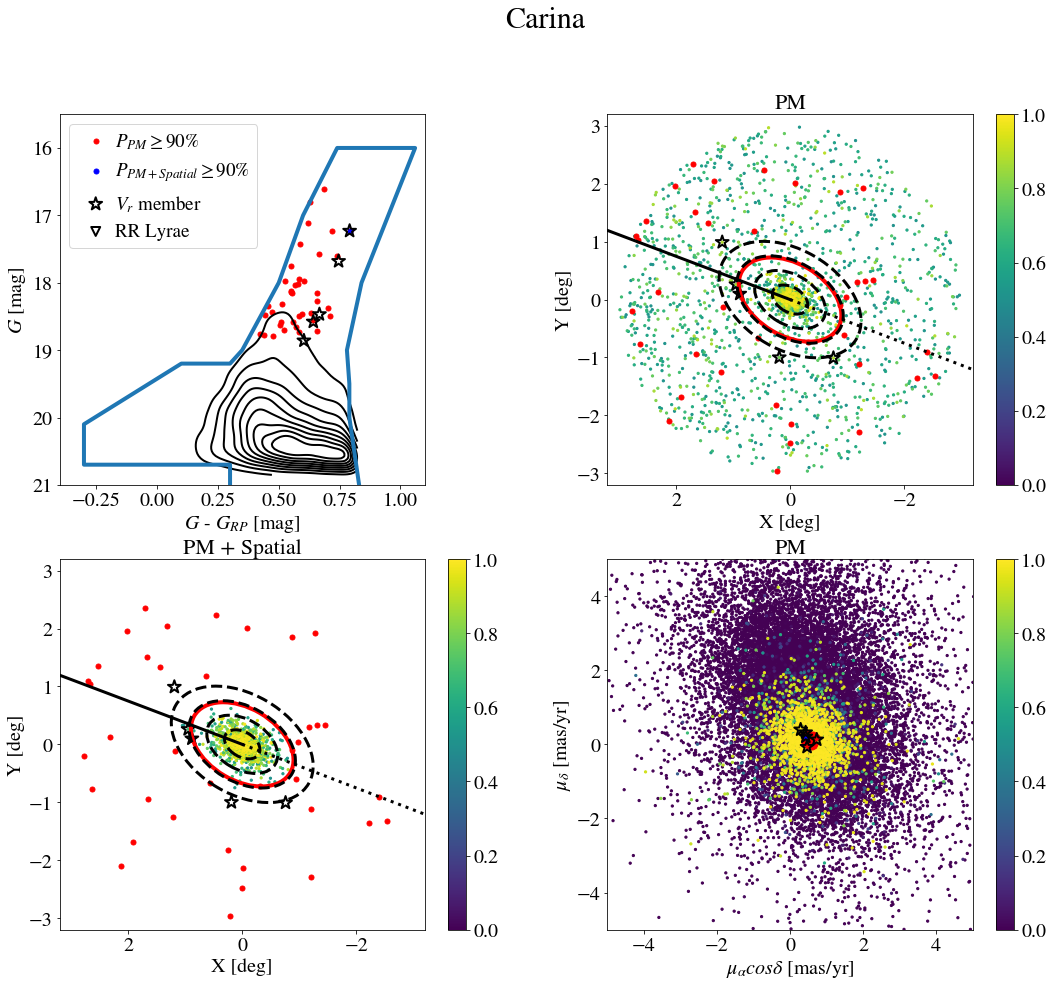} 
  \caption{Same format as Figure \ref{fig:Draco} for Carina. }
  \label{fig:Carina}
\end{figure*}

Beyond the central region of the dSphs, given recent work suggesting the presence of extended stellar features around ultra-faint MW satellites \citep{Chiti2020ApJ...891....8C}, we are interested in attempting to identify similar extended features around the classical dSph satellites.
To this end, we can use our likelihood analysis to identify stars at large radii that are potential members of each dSph.
To visualize the membership probabilities and mark potential outlying members, we set up a series of four plots for each dSph (Figures~\ref{fig:Draco},~\ref{fig:Ursa_Minor},~\ref{fig:Sextans},~\ref{fig:Sculptor},~\ref{fig:Fornax},~\ref{fig:Carina} for Draco, Ursa Minor, Sextans, Sculptor, Fornax, and Carina respectively, hereafter referred to as the membership Figures).
The top right and bottom left plots show the spatial distributions of all stars with $P \geq 50\%$, where the top right shows $P_{PM}$ and the bottom left shows $P_\text{PM+Spatial}$, in order to demonstrate the strong constraints placed on the membership probabilities by requiring a fit to a Plummer profile.
We additionally plot multiples (2, 4, 6, and 8) of the Plummer half-light radius (along with ellipticity and position angle) from  \cite{Munoz18} to provide a quantitative sense of the spatial extent of the dSph.
The bottom right plot shows our full catalog of stars for each dSph field of view, color coded by $P_{PM}$ in PM space.
The top left plot shows the CMD selection used for each dSph, the density of likely MW foreground/background stars in the field of view (stars with $P_{PM} < 1\%$), and stars at large radii that are candidate member stars.

To discuss stars at large radii, we select a boundary to differentiate stars considered to be part of the main body and stars that comprise any extended stellar feature.
For this boundary, given its common usage in discussing the tidal limit of resolved stellar populations, we choose to use the King limiting radius, $r_{t, k}$, position angle, and ellipticity from dedicated photometric work (see Table \ref{tab:properties_lit} for specific values).
In the top right and bottom left plots in the membership Figures, we display this boundary as a red ellipse and mark in red any stars that sit beyond this boundary and have $P_{PM} \geq 90\%$, labeling them as extra-tidal candidates (which are then also displayed in the upper left CMD plot and the bottom right PM plot in the membership Figures).

The locations of the extra-tidal candidates in CMD space may be compared against the MW foreground/background contours as a way to assess potential MW contamination.
For several systems (Draco, Ursa Minor, Sextans, and Carina) we see a clear separation between our high-probability members and the MW population. 
This provides us with strong hints that these stars are indeed associated with the dSphs, even though we have not directly included any information about colors or magnitudes in the likelihood analysis. 
The properties of all extra-tidal candidates (with dust-corrected $G < 19.0$) are shown in Table~\ref{tab:Member} $\sim$ \ref{tab:Member_C3} (a table with all extra-tidal candidates down to our limiting magnitude of $G \sim 20.8$ is available in machine-readable format).

As a first initial check on what kind of contamination we might expect, we test our methodology on a simulated dwarf galaxy orbiting a Milky Way-like host. In particular, we use the Aurigaia catalogs \citep{Grand18} which are mock observations, based on Gaia DR2 systematics, of select Auriga simulations \citep{Grand17}. After applying the same cuts and processing the data with our MCMC framework, we find that $0.16\%$ of the nearby ``Milky Way'' stars pass the $P_{PM} \geq 90\%$ cut and $2.35\%$ pass the $P_{PM} \geq 50\%$ cut. To translate that to our specific dSphs examined we can use the full EDR3 sample as an expected upper limit. In the case of Draco with 3226 stars, we would expect 5.05 stars pass the $P_{PM} \geq 90\%$ cut and 75.73 stars pass the $P_{PM} \geq 50\%$ cut. When we compare that with our identified extra-tidal candidate stars for Draco (18 stars with $P_{PM} \geq 90\%$ and 167 stars with $P_{PM} \geq 50\%$), we see this suggests our extra-tidal candidates to likely be a real feature.

\subsection{Cross match with RR Lyrae}
Given this plausible extra-tidal signal, we want to independently assess the membership of these stars as a check on our PM-only probabilities  ($P_{PM}$).
One way of determining whether a star belongs to a given satellite would be using distance estimates, like the ones provided for variable stars.
In particular, RR Lyrae serve as an excellent map of substructure in the halo, so we can search for RR Lyrae near the center of each satellite that have magnitudes within 0.3 dex in $G$ (corresponding to a distance tolerance of $\sim 10$ kpc for our closer satellites) from the Horizontal Branch (HB) of that satellite.

For this, we compile a collection of variable star catalogs and cross-match each one with all stars in the \gaia\ EDR3 catalog (with no cuts applied) for each satellite field of view.
We combine each cross-matched catalog together, using the \gaia\ EDR3 \texttt{source\_id} column to create a unique list of RR Lyrae.
We then use the \gaia\ magnitudes to select all RR Lyrae within 0.3 dex of the HB for the satellite.
Finally we compare the final positions of these HB-consistent RR Lyrae against the King limiting radius, $r_{t, k}$, for each satellite.
As discussed in Section \ref{sec:data}, we make a series of cuts to select astrometically well-behaved stars, some of which preferentially remove variable stars.
As such, for this RR Lyrae check, we hold off on cross-matching the RR Lyrae with our astrometric catalog until the final step to check against our probabilities.
Below, we list the number of RR Lyrae we find in \gaia\ EDR3 for each satellite, the full list of catalogs we find matching RR Lyrae in, the number that are consistent in distance with the satellite, the number of those outside $r_{t, k}$, and of the extra-tidal RR Lyrae, the number that have a match in our $P_{PM} \geq 50\%$ catalog.


\defcitealias{Vivas04}{V04}
\defcitealias{Kinemuchi08}{K08}
\defcitealias{Drake13a}{D13a}
\defcitealias{Drake13b}{D13b}
\defcitealias{Palaversa13}{P13}
\defcitealias{Vivas13}{V13}
\defcitealias{Drake14}{D14}
\defcitealias{Torrealba15}{T15}
\defcitealias{Drake17}{D17}
\defcitealias{Clementini19}{C19}
\defcitealias{Stringer21}{S21}

\textbf{Draco:} We find 357 unique RR Lyrae (\citealt{Kinemuchi08, Drake13a, Drake13b, Palaversa13, Drake14, Samus17, Sesar17, Clementini19}) in our full EDR3 catalog. 
From those 357 RR Lyrae, 286 have a magnitude consistent with the distance of Draco with 10 sitting outside $r_{t, k}$. 
2 of these 10 are present in our astrometrically-cleaned catalog and have $P_{PM} \geq 50\%$ (as marked in Figure \ref{fig:Draco}).

\textbf{Ursa Minor:} We find 175 unique RR Lyrae (\citealt{Palaversa13, Drake14, Sesar17, Clementini19})
in our full EDR3 catalog. 
From those 175 RR Lyrae, 128 have a magnitude consistent with the distance of Ursa Minor with 12 sitting outside $r_{t, k}$. 
9 of these 12 are present in our catalog and have $P_{PM} \geq 50\%$ (as marked in Figure \ref{fig:Ursa_Minor}).

\textbf{Sextans:} We find 249 unique RR Lyrae
(\citealt{Vivas04, Drake13a, Drake13b, Palaversa13, Drake14, Samus17, Sesar17, Vivas19, Clementini19}) in our full EDR3 catalog.
We do note the presence of the globular cluster Palomar 3 in our field of view and mask out the area around the cluster, which removes 9 RR Lyrae for a total of 274 RR Lyrae remaining. 
Of those 274, 222 have a magnitude consistent with the distance of Sextans with 21 sitting outside $r_{t, k}$. 7 of those 21 are present in our catalog and have $P_{PM} \geq 50\%$ (as marked in Figure \ref{fig:Sextans}).

\textbf{Sculptor:} We find 482 unique RR Lyrae
(\citealt{Torrealba15, Drake17, Clementini19, Stringer21}) 
in our full EDR3 catalog. From those 482 RR Lyrae, 456 have a magnitude consistent with the distance of Sculptor with 2 sitting outside $r_{t, k}$.
1 of these 2 are present in our catalog and have $P_{PM} \geq 50\%$ (as marked in Figure \ref{fig:Sculptor}).

\textbf{Fornax:} We find 67 unique RR Lyrae
(\citealt{Torrealba15, Drake17, Clementini19, Stringer21})
in our full EDR3 catalog. From those 67, 31 have a magnitude consistent with the distance of Fornax with 4 sitting outside $r_{t, k}$. 
1 of these 4 is present in our catalog and has $P_{PM} \geq 50\%$ (as marked in Figure \ref{fig:Fornax}).

\textbf{Carina:} We find 82 unique RR Lyrae
(\citealt{Vivas13, Torrealba15, Drake17, Clementini19})
in our full EDR3 catalog. From those 82 RR Lyrae, 44 have magnitudes consistent with the distance of Carina with 1 sitting outside $r_{t, k}$ but that star does not have $P_{PM} \geq 50\%$, so there are no such joint candidates for Carina.

As a general note, we do not claim original detections of extratidal RR Lyrae for most of the satellites above.
Nearly all of the satellites in this paper have had dedicated work in examining their RR Lyrae populations.
And indeed, many of them note the presence of RR Lyrae consistent with the satellite that lay outside the tidal radius.
For two of these satellites, Fornax and Carina, we do not attempt a rigorous comparison to the literature (\citealt{Vivas13} for Carina and \citealt{Stringer21} for Fornax) as their distance places a significant fraction of the RR Lyrae in the satellite at a magnitude fainter than our \gaia\ magnitude limit.
However, even with our limited sample, we find qualitative agreement between our work and the literature in the distribution of RR Lyrae, with Carina being mostly tightly clustered within the tidal radius and Fornax spilling over the tidal radius.

With respect to the other satellites, we find good agreement with published results for Draco \citep{Muraveva20}, including the extension in the southwest direction in the RR Lyrae population.
Intriguingly, they did not examine too far beyond the King limiting radius for Draco, so our two most outlying PM-consistent RR Lyrae candidate members are absent from their analysis.
For both Sextans \citep{Medina18, Vivas19} and Sculptor \citep{MV16} we also recover the extended structure present in them.
Ursa Minor is the only one of the six without a publicly available dedicated analysis, so we were unable to compare our results.
Taken as a whole, the identification of a majority of RR Lyrae in our PM membership sample, despite the astrometric quality cuts, the consistency with published results, and still finding matches to RR Lyrae in the extratidal candidates, is encouraging evidence for the robustness of our methodology.

\defcitealias{Simon19}{S19}
\defcitealias{Walker15}{W15}
\defcitealias{Walker09}{W09}
\defcitealias{Spencer18}{Sp18}
\defcitealias{Battaglia11}{B11}

\subsection{Cross match with radial velocities}
\label{sec:RV}

\par For our sample of high-probability stars, we search the literature for stars that have existing radial velocities. If a given star does indeed have RV consistent with the bulk radial velocity of the dSph, it would give us both high confidence in our analysis method, as well as a confirmation that the star is an associated member of the dSph. 

We compile a collection of RV catalogs and cross-match each one with our \gaia EDR3 catalog (see Section \ref{sec:data}). We then keep only stars with consistent RV (within $3\sigma_{v}$ of $V_r$, Table \ref{tab:properties_lit}) in our RV catalogs. In our consistent RV catalogs, we define a star with $P_{PM} \geq 50\%$ as a $V_r$ member. Below, we list the number of RV matches we find in our full \gaia EDR3 catalog for each satellite, the full list of catalogs we find matching RV in, the number that are consistent in the bulk RV, and of the RV beyond King limiting radius, the number that have $P_{PM} \geq 50\% $, and bright (dust-corrected $G\leq 19$) stars  with $P_{PM} \geq 90\% $.

\textbf{Draco:} We find 636 stars with RV data 
(\citealt{Walker15, Spencer18})
in our full EDR3 catalog. 
From those 505 have RV that are within $3\sigma_{v}$ of the RV as obtained from \cite{Simon19}, hereafter \citetalias{Simon19} (see Table \ref{tab:properties_lit}). 
From those 505 stars, 425 have $P_{PM} \geq 50\%$, with 4 sitting outside $r_{t, k}$ (see $V_r$ member in Figure \ref{fig:Draco}), suggesting more candidates may be identified with considering stars with lower membership probabilities. 
Out of the 10 brightest ($G \leq 19$) possible candidates ($P_{PM} \geq 90\%$) beyond the King limiting radius in Table \ref{tab:Member}, we find 2 stars with matching radial velocities. These two stars are labeled with red stars in Figure \ref{fig:Draco}.  

\textbf{Ursa Minor:} We find 762 stars with RV data
(\citealt{Spencer18, Pace20})
in our full EDR3 catalog. From those 731 have RV that are within $3\sigma_{v}$ of the RV as obtained from \citetalias{Simon19}. From those 731 stars, 665 have $P_{PM} \geq 50\%$, with 2 sitting outside $r_{t, k}$ (Figure \ref{fig:Ursa_Minor}).
Out of the 30 brightest ($G \leq 19$) possible candidates in Table \ref{tab:Member}, we find 1 star with matching radial velocities. 

\textbf{Sextans:} We find 540 stars with RV data 
(\citealt{Walker09, Walker15, Battaglia11})
in our full EDR3 catalog. From those 478 have RV that are within $3\sigma_{v}$ of the RV as obtained from \citetalias{Simon19}. Among them, 412 have $\geq 50\%$ PM-only membership probability, with 8 sitting outside $r_{t, k}$ (Figure  \ref{fig:Sextans}).
Out of the 34 brightest possible candidates in Table \ref{tab:Member_C1}, we find 4 stars with matching radial velocities.

\textbf{Sculptor:} We find 1423 stars with RV data 
(\citealt{Walker09, Walker15})
in our full EDR3 catalog. From those 1370 have RV that are within $3\sigma_{v}$ of the RV as obtained from \citetalias{Simon19}. Among them, 1355 have $\geq 50\%$ PM-only membership probability, with 2 sitting outside $r_{t, k}$ (Figure \ref{fig:Sculptor}).
Out of the 10 brightest possible candidates in Table \ref{tab:Member_C1}, we find 2 stars with matching radial velocities.

\textbf{Fornax:} We find 2721 stars with RV data 
(\citealt{Walker09, Hendricks14, Pace21})
in our full EDR3 catalog. From those 2667 have RV that are within $3\sigma_{v}$ of the RV as obtained from \citetalias{Simon19}. Among them, 2662 have $\geq 50\%$ PM-only membership probability, with 12 sitting outside $r_{t, k}$ (Figure \ref{fig:Fornax}). 
Out of the 30 brightest possible candidates in Table \ref{tab:Member_C2}, we find 5 stars with matching radial velocities. 

\textbf{Carina:} We find 1443 stars with RV data 
(\citealt{Munoz06, Walker09, Fabrizio16})
in our full EDR3 catalog. From those 1190 have RV that are within $3\sigma_{v}$ of the RV as obtained from \citetalias{Simon19}. Among them, 717 have $\geq 50\%$ PM-only membership probability, with 5 sitting outside $r_{t, k}$ (Figure \ref{fig:Carina}).
Out of the 39 brightest possible candidates in Table \ref{tab:Member_C3}, we find 1 star with matching radial velocities. 

\subsection{Orbits} 

Tidal debris from disrupting satellites generally aligns spatially with the orbit of the progenitor \citep{Dehnen:2004, Montuori:2007}, a fact that can be used to associate stellar streams with previously-known objects \citep{Ibata:2019, Hansen:2020, Bonaca:2021}.
We compute orbits for each satellite using \texttt{gala} \citep{gala} to see if our high-probability candidates are found near the orbital track.
We represent the potential of the Milky Way using a four-component model composed of a spherical Hernquist bulge \citep{Hernquist:1990}, a spherical Hernquist nucleus, an axisymmetric Miyamoto-Nagai disk \citep{Miyamoto:1975}, and a spherical Navarro-Frenk-White dark matter halo \citep{Navarro:1996}, with parameters fixed to their default values in \texttt{gala} \citep[v1.3;][]{gala-v1.3}.
We assume the on-sky coordinates, distance, and radial velocity for each satellite as compiled in Table \ref{tab:properties_lit} and the proper motions derived in this work from Table \ref{tab:Properties}.
To convert these heliocentric observations to a Galactocentric frame, we adopt a distance from the Sun to the Galactic center of $R_0 = 8.122$~kpc \citep{GravityCollaboration:2018}, a height of the Sun relative to the Galactic plane of $z_0 = 20.8$~pc \citep{Bennett:2019}, and a solar motion relative to the Galactic center of $(12.9, 245.6, 7.78)$~km~s$^{-1}$ \citep{Drimmel:2018, GravityCollaboration:2018, Reid:2004}.

In general, our extratidal PM-only candidates ($P_{PM} \geq 90 \%$) are distributed roughly uniform along the sky, with some candidates lying very close to the orbital track.
PM+spatial candidates ($P_\text{PM+Spatial} \geq 90 \%$) tend to lie within one degree of the orbit track, but this is largely expected due to the spatial component of the likelihood penalizing large on-sky separations from the satellite, which lies (naturally) along the orbit.
We particularly highlight Sculptor, where two of the four PM+spatial candidates and the confirmed radial velocity member lie very close to the trailing orbit, as well as Draco, whose one radial velocity member is within one degree of the leading orbit.
Extratidal candidates that lie close to the satellite's orbital track but far from the satellite itself are promising targets for spectroscopic followup.

\section{Discussion and Conclusion}
\label{sec:conclusions}
\par We have used \gaia EDR3 astrometry data to identify stars associated with six classical dwarf spheroidals (Draco, Ursa Minor, Sextans, Sculptor, Fornax, Carina) at their outermost radii. Using a gaussian mixture model to describe the proper motion distributions for the member stars and for the background of each dSph, we identify a substantial population of candidate associations for each dSph. All of our astrometric members are consistent with the color-magnitude diagrams for each dSph. In summary, we identify a list of extra-tidal candidates ($P_{PM} \geq 90\%$) in all six dSphs. We find 18 extra-tidal candidates in Draco, 95 extra-tidal candidates in Ursa Minor, 40 extra-tidal candidates in Sextans, 334 extra-tidal candidates in Sculptor, 844 extra-tidal candidates in Fornax, and 39 extra-tidal candidates in Carina.
We use simulation to calculate that $0.16\%$ of the Milky Way stars pass the $P_{PM} \geq 90\%$ cut, and suggest that our extra-tidal candidates to likely be a real feature.

Since our methodology for assigning membership probabilities does not include any information on radial velocity (RV) or distance, when cross-matching with RV samples and RR Lyrae catalogs in the literature,  we use $P_{PM} \geq 50\%$ stars to include more matches.

Using this cross-match, we identify a list of candidates outside the King limiting radius in all six dSphs consisent with both RV and astrometry data. From this radial velocity matching, we find 4 candidates in Draco, 2 in Ursa Minor, 8 in Sextans, 2 in Sculptor, 12 in Fornax and 5 in Carina, indicating that these stars are associated with their respective dSphs at high probability. This provides strong kinematic evidence for the existence of member stars associated with the dSphs outside of their classical stellar King limiting radius. 

\par In order to further improve membership assessment, we have additionally cross-matched our samples of  candidates to RR Lyrae catalogs. Restricting to stars with astrometric membership probability of $P_{PM} \geq 50\%$, we find 1 RR Lyrae candidate outside the limiting radius in 
both Sculptor and Fornax. We find 2 such candidate for Draco, 9 for Ursa Minor, and 7 for Sextans. No matches with our $P_{PM} \geq 50\%$ catalog were found outside the limiting radius for Carina.

\par At this stage, we are not able to determine whether the aforementioned candidates represent an extended stellar halo component, or possibly even tidal debris that has been removed from the main body of the dSph due to interactions with the Milky Way gravitational potential. If the candidate stars that we identify above are associated with tidal debris from the dSphs, it is may be the case that these candidate stars line up along the orbital track of the dSph. Using updated estimates for the orbital tracks for each dSph, and models for the potential of the Milky Way dark matter halo, we compare the projected spatial distribution of the star candidates to the best-fitting orbital tracks for each dSph. In the case of Sculptor, we find possible alignment between the orbital tracks and our stellar candidates, while for the remaining dSphs the stellar candidates are more or less randomly distributed relative to the orbital tracks. Further precision measurements of the orbits of dSphs and firm identification of member stars at large radii will improve the prospects for identifying faint tidal debris around dSphs. 

\par The candidate stars that we have identified, both those candidates that are associated with RR Lyrae and those that are not, provide an optimal sample of candidate stars for spectroscopic follow-up. Radial velocities and metallicities can be used to confirm the membership of our distant candidates. In particular for the RR Lyrae member candidates, both astrometry and radial velocities would complete a full 6D phase-space coverage on the kinematic properties of these stars. 

\par Though current~\gaia~data is not sensitive to the measurement of internal tangential velocity dispersions of dSphs ($\sim$10 km/s), it is possible that future~\gaia~data releases will come close to, or measure, these dispersions. For this reason for the analysis in this paper, we have fixed the internal velocity dispersions of the dSphs. However, the formalism that we have implemented in this paper may be naturally extended to measure or places bounds on the tangential dispersions of dSphs with forthcoming data. 
In addition, the most distant stars are ideal for constraining potential rotation or velocity gradients. 

\par Though in this paper we have assumed gaussian models for the velocity distributions of the foreground and background, the modeling in this paper may ultimately be improved upon by constructing alternative models for the velocity distributions, in particular for that of the foreground. The formalism we have presented may be adapted to numerical simulations of the distribution which generate a population of dSphs in a Milky Way-like halo, which includes stellar halo components. Examining such simulated halos would both provide a sense for how the foregrounds may differ from the gaussian model that we have assumed, and provide a sense of the biases that are incurred by using a gaussian model as we have implemented. 

\section{Acknowledgements}
LES acknowledges support from DOE Grant de-sc0010813.  
This work was supported by a Development Fellowship from the Texas A$\&$M University System National Laboratories Office. 
ABP is supported by NSF grant AST-1813881.
AHR acknowledges support from an NSF Graduate Research Fellowship through Grant DGE-1746932. We are grateful to Rob Grand for providing us data on the properties of dwarf galaxies in the Auriga simulation.This work made use of Python Programming Language, and software packages including \texttt{Jupyter} \citep{Jupyter}, \texttt{Astropy} \citep{Astropy13,Astropy18}, \texttt{galpy} \citep{galpy15}, \texttt{numpy} \citep{NumPy20}, \texttt{scipy} \citep{SciPy20}, and \texttt{corner.py} \citep{cornerpy16}. This research has made use of NASA’s Astrophysics
Data System for bibliographic services.

This work presents results from the European Space Agency (ESA) space mission \gaia. \gaia data are being processed by the Gaia Data Processing and Analysis Consortium (DPAC). Funding for the DPAC is provided by national institutions, in particular the institutions participating in the \gaia MultiLateral Agreement (MLA). The \gaia mission website is https://www.cosmos.esa.int/gaia. The \gaia archive website is https://archives.esac.esa.int/gaia.

\section{Data Availability}
The data underlying this article will be shared on reasonable request to the corresponding author. 

\bibliographystyle{mnras}
\bibliography{main}

\begin{thebibliography}{}
\makeatletter
\relax
\def\mn@urlcharsother{\let\do\@makeother \do\$\do\&\do\#\do\^\do\_\do\%\do\~}
\def\mn@doi{\begingroup\mn@urlcharsother \@ifnextchar [ {\mn@doi@}
  {\mn@doi@[]}}
\def\mn@doi@[#1]#2{\def\@tempa{#1}\ifx\@tempa\@empty \href
  {http://dx.doi.org/#2} {doi:#2}\else \href {http://dx.doi.org/#2} {#1}\fi
  \endgroup}
\def\mn@eprint#1#2{\mn@eprint@#1:#2::\@nil}
\def\mn@eprint@arXiv#1{\href {http://arxiv.org/abs/#1} {{\tt arXiv:#1}}}
\def\mn@eprint@dblp#1{\href {http://dblp.uni-trier.de/rec/bibtex/#1.xml}
  {dblp:#1}}
\def\mn@eprint@#1:#2:#3:#4\@nil{\def\@tempa {#1}\def\@tempb {#2}\def\@tempc
  {#3}\ifx \@tempc \@empty \let \@tempc \@tempb \let \@tempb \@tempa \fi \ifx
  \@tempb \@empty \def\@tempb {arXiv}\fi \@ifundefined
  {mn@eprint@\@tempb}{\@tempb:\@tempc}{\expandafter \expandafter \csname
  mn@eprint@\@tempb\endcsname \expandafter{\@tempc}}}

\bibitem[\protect\citeauthoryear{{Astropy Collaboration} et~al.,}{{Astropy
  Collaboration} et~al.}{2013}]{Astropy13}
{Astropy Collaboration} et~al., 2013, \mn@doi [\aap]
  {10.1051/0004-6361/201322068}, \href
  {https://ui.adsabs.harvard.edu/abs/2013A&A...558A..33A} {558, A33}

\bibitem[\protect\citeauthoryear{{Astropy Collaboration} et~al.,}{{Astropy
  Collaboration} et~al.}{2018}]{Astropy18}
{Astropy Collaboration} et~al., 2018, \mn@doi [\aj] {10.3847/1538-3881/aabc4f},
  \href {https://ui.adsabs.harvard.edu/abs/2018AJ....156..123A} {156, 123}

\bibitem[\protect\citeauthoryear{{Battaglia} et~al.,}{{Battaglia}
  et~al.}{2006}]{Battaglia06}
{Battaglia} G.,  et~al., 2006, \mn@doi [\aap] {10.1051/0004-6361:20065720},
  \href {https://ui.adsabs.harvard.edu/abs/2006A&A...459..423B} {459, 423}

\bibitem[\protect\citeauthoryear{{Battaglia}, {Tolstoy}, {Helmi}, {Irwin},
  {Parisi}, {Hill}  \& {Jablonka}}{{Battaglia} et~al.}{2011}]{Battaglia11}
{Battaglia} G.,  {Tolstoy} E.,  {Helmi} A.,  {Irwin} M.,  {Parisi} P.,  {Hill}
  V.,   {Jablonka} P.,  2011, \mn@doi [\mnras]
  {10.1111/j.1365-2966.2010.17745.x}, \href
  {https://ui.adsabs.harvard.edu/abs/2011MNRAS.411.1013B} {411, 1013}

\bibitem[\protect\citeauthoryear{{Battaglia}, {Helmi}  \&
  {Breddels}}{{Battaglia} et~al.}{2013a}]{2013NewAR..57...52B}
{Battaglia} G.,  {Helmi} A.,   {Breddels} M.,  2013a, \mn@doi [\nar]
  {10.1016/j.newar.2013.05.003}, \href
  {https://ui.adsabs.harvard.edu/abs/2013NewAR..57...52B} {57, 52}

\bibitem[\protect\citeauthoryear{{Battaglia}, {Helmi}  \&
  {Breddels}}{{Battaglia} et~al.}{2013b}]{Battaglia:2013}
{Battaglia} G.,  {Helmi} A.,   {Breddels} M.,  2013b, \mn@doi [\nar]
  {10.1016/j.newar.2013.05.003}, \href
  {https://ui.adsabs.harvard.edu/abs/2013NewAR..57...52B} {57, 52}

\bibitem[\protect\citeauthoryear{{Battaglia}, {Sollima}  \&
  {Nipoti}}{{Battaglia} et~al.}{2015}]{2015MNRAS.454.2401B}
{Battaglia} G.,  {Sollima} A.,   {Nipoti} C.,  2015, \mn@doi [\mnras]
  {10.1093/mnras/stv2096}, \href
  {https://ui.adsabs.harvard.edu/abs/2015MNRAS.454.2401B} {454, 2401}

\bibitem[\protect\citeauthoryear{{Battaglia}, {Taibi}, {Thomas}  \&
  {Fritz}}{{Battaglia} et~al.}{2021}]{Battaglia21}
{Battaglia} G.,  {Taibi} S.,  {Thomas} G.~F.,   {Fritz} T.~K.,  2021, arXiv
  e-prints, \href {https://ui.adsabs.harvard.edu/abs/2021arXiv210608819B} {p.
  arXiv:2106.08819}

\bibitem[\protect\citeauthoryear{{Bennett} \& {Bovy}}{{Bennett} \&
  {Bovy}}{2019}]{Bennett:2019}
{Bennett} M.,  {Bovy} J.,  2019, \mn@doi [\mnras] {10.1093/mnras/sty2813},
  \href {https://ui.adsabs.harvard.edu/abs/2019MNRAS.482.1417B} {482, 1417}

\bibitem[\protect\citeauthoryear{{Bonaca} et~al.,}{{Bonaca}
  et~al.}{2021}]{Bonaca:2021}
{Bonaca} A.,  et~al., 2021, \mn@doi [\apjl] {10.3847/2041-8213/abeaa9}, \href
  {https://ui.adsabs.harvard.edu/abs/2021ApJ...909L..26B} {909, L26}

\bibitem[\protect\citeauthoryear{{Bovy}}{{Bovy}}{2015}]{galpy15}
{Bovy} J.,  2015, \mn@doi [\apjs] {10.1088/0067-0049/216/2/29}, \href
  {https://ui.adsabs.harvard.edu/abs/2015ApJS..216...29B} {216, 29}

\bibitem[\protect\citeauthoryear{{Bullock} \& {Boylan-Kolchin}}{{Bullock} \&
  {Boylan-Kolchin}}{2017}]{2017ARA&A..55..343B}
{Bullock} J.~S.,  {Boylan-Kolchin} M.,  2017, \mn@doi [\araa]
  {10.1146/annurev-astro-091916-055313}, \href
  {https://ui.adsabs.harvard.edu/abs/2017ARA&A..55..343B} {55, 343}

\bibitem[\protect\citeauthoryear{{Carlin} \& {Sand}}{{Carlin} \&
  {Sand}}{2018}]{Carlin18}
{Carlin} J.~L.,  {Sand} D.~J.,  2018, \mn@doi [\apj]
  {10.3847/1538-4357/aad8c1}, \href
  {https://ui.adsabs.harvard.edu/abs/2018ApJ...865....7C} {865, 7}

\bibitem[\protect\citeauthoryear{{Chiti}, {Frebel}, {Jerjen}, {Kim}  \&
  {Norris}}{{Chiti} et~al.}{2020}]{Chiti2020ApJ...891....8C}
{Chiti} A.,  {Frebel} A.,  {Jerjen} H.,  {Kim} D.,   {Norris} J.~E.,  2020,
  \mn@doi [\apj] {10.3847/1538-4357/ab6d72}, \href
  {https://ui.adsabs.harvard.edu/abs/2020ApJ...891....8C} {891, 8}

\bibitem[\protect\citeauthoryear{{Clementini} et~al.,}{{Clementini}
  et~al.}{2019}]{Clementini19}
{Clementini} G.,  et~al., 2019, \mn@doi [\aap] {10.1051/0004-6361/201833374},
  \href {https://ui.adsabs.harvard.edu/abs/2019A&A...622A..60C} {622, A60}

\bibitem[\protect\citeauthoryear{{Deason}, {Bose}, {Fattahi}, {Amorisco},
  {Hellwing}  \& {Frenk}}{{Deason} et~al.}{2021}]{Deason21}
{Deason} A.~J.,  {Bose} S.,  {Fattahi} A.,  {Amorisco} N.~C.,  {Hellwing} W.,
  {Frenk} C.~S.,  2021, \mn@doi [\mnras] {10.1093/mnras/stab3524}, \href
  {https://ui.adsabs.harvard.edu/abs/2021MNRAS.tmp.3180D} {}

\bibitem[\protect\citeauthoryear{{Dehnen}, {Odenkirchen}, {Grebel}  \&
  {Rix}}{{Dehnen} et~al.}{2004}]{Dehnen:2004}
{Dehnen} W.,  {Odenkirchen} M.,  {Grebel} E.~K.,   {Rix} H.-W.,  2004, \mn@doi
  [\aj] {10.1086/383214}, \href
  {https://ui.adsabs.harvard.edu/abs/2004AJ....127.2753D} {127, 2753}

\bibitem[\protect\citeauthoryear{{Drake} et~al.,}{{Drake}
  et~al.}{2013a}]{Drake13a}
{Drake} A.~J.,  et~al., 2013a, \mn@doi [\apj] {10.1088/0004-637X/763/1/32},
  \href {https://ui.adsabs.harvard.edu/abs/2013ApJ...763...32D} {763, 32}

\bibitem[\protect\citeauthoryear{{Drake} et~al.,}{{Drake}
  et~al.}{2013b}]{Drake13b}
{Drake} A.~J.,  et~al., 2013b, \mn@doi [\apj] {10.1088/0004-637X/765/2/154},
  \href {https://ui.adsabs.harvard.edu/abs/2013ApJ...765..154D} {765, 154}

\bibitem[\protect\citeauthoryear{{Drake} et~al.,}{{Drake}
  et~al.}{2014}]{Drake14}
{Drake} A.~J.,  et~al., 2014, \mn@doi [\apjs] {10.1088/0067-0049/213/1/9},
  \href {https://ui.adsabs.harvard.edu/abs/2014ApJS..213....9D} {213, 9}

\bibitem[\protect\citeauthoryear{{Drake} et~al.,}{{Drake}
  et~al.}{2017}]{Drake17}
{Drake} A.~J.,  et~al., 2017, \mn@doi [\mnras] {10.1093/mnras/stx1085}, \href
  {https://ui.adsabs.harvard.edu/abs/2017MNRAS.469.3688D} {469, 3688}

\bibitem[\protect\citeauthoryear{{Drimmel} \& {Poggio}}{{Drimmel} \&
  {Poggio}}{2018}]{Drimmel:2018}
{Drimmel} R.,  {Poggio} E.,  2018, \mn@doi [Research Notes of the American
  Astronomical Society] {10.3847/2515-5172/aaef8b}, \href
  {https://ui.adsabs.harvard.edu/abs/2018RNAAS...2..210D} {2, 210}

\bibitem[\protect\citeauthoryear{{Drlica-Wagner} et~al.,}{{Drlica-Wagner}
  et~al.}{2015}]{Drlica-Wagner15}
{Drlica-Wagner} A.,  et~al., 2015, \mn@doi [\apj]
  {10.1088/0004-637X/813/2/109}, \href
  {https://ui.adsabs.harvard.edu/abs/2015ApJ...813..109D} {813, 109}

\bibitem[\protect\citeauthoryear{{Fabrizio} et~al.,}{{Fabrizio}
  et~al.}{2016}]{Fabrizio16}
{Fabrizio} M.,  et~al., 2016, \mn@doi [\apj] {10.3847/0004-637X/830/2/126},
  \href {https://ui.adsabs.harvard.edu/abs/2016ApJ...830..126F} {830, 126}

\bibitem[\protect\citeauthoryear{{Fillingham} et~al.,}{{Fillingham}
  et~al.}{2019}]{Fillingham19}
{Fillingham} S.~P.,  et~al., 2019, arXiv e-prints, \href
  {https://ui.adsabs.harvard.edu/abs/2019arXiv190604180F} {p. arXiv:1906.04180}

\bibitem[\protect\citeauthoryear{{Foreman-Mackey}}{{Foreman-Mackey}}{2016}]{cornerpy16}
{Foreman-Mackey} D.,  2016, \mn@doi [The Journal of Open Source Software]
  {10.21105/joss.00024}, \href
  {https://ui.adsabs.harvard.edu/abs/2016JOSS....1...24F} {1, 24}

\bibitem[\protect\citeauthoryear{{Foreman-Mackey}, {Hogg}, {Lang}  \&
  {Goodman}}{{Foreman-Mackey} et~al.}{2013}]{Foreman-Mackey13}
{Foreman-Mackey} D.,  {Hogg} D.~W.,  {Lang} D.,   {Goodman} J.,  2013, \mn@doi
  [\pasp] {10.1086/670067}, \href
  {https://ui.adsabs.harvard.edu/abs/2013PASP..125..306F} {125, 306}

\bibitem[\protect\citeauthoryear{{Fritz}, {Battaglia}, {Pawlowski},
  {Kallivayalil}, {van der Marel}, {Sohn}, {Brook}  \& {Besla}}{{Fritz}
  et~al.}{2018}]{Fritz18}
{Fritz} T.~K.,  {Battaglia} G.,  {Pawlowski} M.~S.,  {Kallivayalil} N.,  {van
  der Marel} R.,  {Sohn} S.~T.,  {Brook} C.,   {Besla} G.,  2018, \mn@doi
  [\aap] {10.1051/0004-6361/201833343}, \href
  {https://ui.adsabs.harvard.edu/abs/2018A&A...619A.103F} {619, A103}

\bibitem[\protect\citeauthoryear{{Gaia Collaboration} et~al.,}{{Gaia
  Collaboration} et~al.}{2016}]{Gaia16}
{Gaia Collaboration} et~al., 2016, \mn@doi [\aap]
  {10.1051/0004-6361/201629272}, \href
  {https://ui.adsabs.harvard.edu/abs/2016A&A...595A...1G} {595, A1}

\bibitem[\protect\citeauthoryear{{Gaia Collaboration} et~al.,}{{Gaia
  Collaboration} et~al.}{2018a}]{Babusiaux18}
{Gaia Collaboration} et~al., 2018a, \mn@doi [\aap]
  {10.1051/0004-6361/201832843}, \href
  {https://ui.adsabs.harvard.edu/abs/2018A&A...616A..10G} {616, A10}

\bibitem[\protect\citeauthoryear{{Gaia Collaboration} et~al.,}{{Gaia
  Collaboration} et~al.}{2018b}]{Gaia18}
{Gaia Collaboration} et~al., 2018b, \mn@doi [\aap]
  {10.1051/0004-6361/201832698}, \href
  {https://ui.adsabs.harvard.edu/abs/2018A&A...616A..12G} {616, A12}

\bibitem[\protect\citeauthoryear{{Gaia Collaboration}, {Brown}, {Vallenari},
  {Prusti}, {de Bruijne}, {Babusiaux}  \& {Biermann}}{{Gaia Collaboration}
  et~al.}{2020}]{Gaia20}
{Gaia Collaboration} {Brown} A.~G.~A.,  {Vallenari} A.,  {Prusti} T.,  {de
  Bruijne} J.~H.~J.,  {Babusiaux} C.,   {Biermann} M.,  2020, arXiv e-prints,
  \href {https://ui.adsabs.harvard.edu/abs/2020arXiv201201533G} {p.
  arXiv:2012.01533}

\bibitem[\protect\citeauthoryear{{Genina}, {Read}, {Fattahi}  \&
  {Frenk}}{{Genina} et~al.}{2020}]{Genina20}
{Genina} A.,  {Read} J.~I.,  {Fattahi} A.,   {Frenk} C.~S.,  2020, arXiv
  e-prints, \href {https://ui.adsabs.harvard.edu/abs/2020arXiv201109482G} {p.
  arXiv:2011.09482}

\bibitem[\protect\citeauthoryear{{Goodman} \& {Weare}}{{Goodman} \&
  {Weare}}{2010}]{Goodman&Weare10}
{Goodman} J.,  {Weare} J.,  2010, \mn@doi [Communications in Applied
  Mathematics and Computational Science] {10.2140/camcos.2010.5.65}, \href
  {https://ui.adsabs.harvard.edu/abs/2010CAMCS...5...65G} {5, 65}

\bibitem[\protect\citeauthoryear{{Grand} et~al.,}{{Grand}
  et~al.}{2017}]{Grand17}
{Grand} R. J.~J.,  et~al., 2017, \mn@doi [\mnras] {10.1093/mnras/stx071}, \href
  {https://ui.adsabs.harvard.edu/abs/2017MNRAS.467..179G} {467, 179}

\bibitem[\protect\citeauthoryear{{Grand} et~al.,}{{Grand}
  et~al.}{2018}]{Grand18}
{Grand} R. J.~J.,  et~al., 2018, \mn@doi [\mnras] {10.1093/mnras/sty2403},
  \href {https://ui.adsabs.harvard.edu/abs/2018MNRAS.481.1726G} {481, 1726}

\bibitem[\protect\citeauthoryear{{Gravity Collaboration} et~al.,}{{Gravity
  Collaboration} et~al.}{2018}]{GravityCollaboration:2018}
{Gravity Collaboration} et~al., 2018, \mn@doi [\aap]
  {10.1051/0004-6361/201833718}, \href
  {https://ui.adsabs.harvard.edu/abs/2018A&A...615L..15G} {615, L15}

\bibitem[\protect\citeauthoryear{{Gregory} et~al.,}{{Gregory}
  et~al.}{2020}]{2020MNRAS.496.1092G}
{Gregory} A.~L.,  et~al., 2020, \mn@doi [\mnras] {10.1093/mnras/staa1553},
  \href {https://ui.adsabs.harvard.edu/abs/2020MNRAS.496.1092G} {496, 1092}

\bibitem[\protect\citeauthoryear{{Hansen}, {Riley}, {Strigari}, {Marshall},
  {Ferguson}, {Zepeda}  \& {Sneden}}{{Hansen} et~al.}{2020}]{Hansen:2020}
{Hansen} T.~T.,  {Riley} A.~H.,  {Strigari} L.~E.,  {Marshall} J.~L.,
  {Ferguson} P.~S.,  {Zepeda} J.,   {Sneden} C.,  2020, \mn@doi [\apj]
  {10.3847/1538-4357/ababa5}, \href
  {https://ui.adsabs.harvard.edu/abs/2020ApJ...901...23H} {901, 23}

\bibitem[\protect\citeauthoryear{Harris et~al.,}{Harris et~al.}{2020}]{NumPy20}
Harris C.~R.,  et~al., 2020, \mn@doi [Nature] {10.1038/s41586-020-2649-2}, 585,
  357–362

\bibitem[\protect\citeauthoryear{{Hendricks}, {Koch}, {Walker}, {Johnson},
  {Pe{\~n}arrubia}  \& {Gilmore}}{{Hendricks} et~al.}{2014}]{Hendricks14}
{Hendricks} B.,  {Koch} A.,  {Walker} M.,  {Johnson} C.~I.,  {Pe{\~n}arrubia}
  J.,   {Gilmore} G.,  2014, \mn@doi [\aap] {10.1051/0004-6361/201424645},
  \href {https://ui.adsabs.harvard.edu/abs/2014A&A...572A..82H} {572, A82}

\bibitem[\protect\citeauthoryear{{Hernquist}}{{Hernquist}}{1990}]{Hernquist:1990}
{Hernquist} L.,  1990, \mn@doi [\apj] {10.1086/168845}, \href
  {https://ui.adsabs.harvard.edu/abs/1990ApJ...356..359H} {356, 359}

\bibitem[\protect\citeauthoryear{{Ibata}, {Bellazzini}, {Malhan}, {Martin}  \&
  {Bianchini}}{{Ibata} et~al.}{2019}]{Ibata:2019}
{Ibata} R.~A.,  {Bellazzini} M.,  {Malhan} K.,  {Martin} N.,   {Bianchini} P.,
  2019, \mn@doi [Nature Astronomy] {10.1038/s41550-019-0751-x}, \href
  {https://ui.adsabs.harvard.edu/abs/2019NatAs...3..667I} {3, 667}

\bibitem[\protect\citeauthoryear{{Ibata}, {Bellazzini}, {Thomas}, {Malhan},
  {Martin}, {Famaey}  \& {Siebert}}{{Ibata} et~al.}{2020}]{Ibata20}
{Ibata} R.,  {Bellazzini} M.,  {Thomas} G.,  {Malhan} K.,  {Martin} N.,
  {Famaey} B.,   {Siebert} A.,  2020, \mn@doi [\apjl]
  {10.3847/2041-8213/ab77c7}, \href
  {https://ui.adsabs.harvard.edu/abs/2020ApJ...891L..19I} {891, L19}

\bibitem[\protect\citeauthoryear{{Kinemuchi}, {Harris}, {Smith}, {Silbermann},
  {Snyder}, {La Cluyz{\'e}}  \& {Clark}}{{Kinemuchi}
  et~al.}{2008}]{Kinemuchi08}
{Kinemuchi} K.,  {Harris} H.~C.,  {Smith} H.~A.,  {Silbermann} N.~A.,  {Snyder}
  L.~A.,  {La Cluyz{\'e}} A.~P.,   {Clark} C.~L.,  2008, \mn@doi [\aj]
  {10.1088/0004-6256/136/5/1921}, \href
  {https://ui.adsabs.harvard.edu/abs/2008AJ....136.1921K} {136, 1921}

\bibitem[\protect\citeauthoryear{Kluyver et~al.,}{Kluyver
  et~al.}{2016}]{Jupyter}
Kluyver T.,  et~al., 2016, in Loizides F.,  Schmidt B.,  eds, Positioning and
  Power in Academic Publishing: Players, Agents and Agendas. pp 87 -- 90

\bibitem[\protect\citeauthoryear{{Koposov} et~al.,}{{Koposov}
  et~al.}{2012}]{Koposov12}
{Koposov} S.~E.,  et~al., 2012, \mn@doi [\apj] {10.1088/0004-637X/750/1/80},
  \href {https://ui.adsabs.harvard.edu/abs/2012ApJ...750...80K} {750, 80}

\bibitem[\protect\citeauthoryear{{Li} et~al.,}{{Li} et~al.}{2018}]{Li18}
{Li} T.~S.,  et~al., 2018, \mn@doi [\apj] {10.3847/1538-4357/aadf91}, \href
  {https://ui.adsabs.harvard.edu/abs/2018ApJ...866...22L} {866, 22}

\bibitem[\protect\citeauthoryear{{Longeard} et~al.,}{{Longeard}
  et~al.}{2021}]{2021arXiv210710849L}
{Longeard} N.,  et~al., 2021, arXiv e-prints, \href
  {https://ui.adsabs.harvard.edu/abs/2021arXiv210710849L} {p. arXiv:2107.10849}

\bibitem[\protect\citeauthoryear{{Majewski}, {Skrutskie}, {Weinberg}  \&
  {Ostheimer}}{{Majewski} et~al.}{2003}]{Majewski03}
{Majewski} S.~R.,  {Skrutskie} M.~F.,  {Weinberg} M.~D.,   {Ostheimer} J.~C.,
  2003, \mn@doi [\apj] {10.1086/379504}, \href
  {https://ui.adsabs.harvard.edu/abs/2003ApJ...599.1082M} {599, 1082}

\bibitem[\protect\citeauthoryear{{Mart{\'\i}nez-Garc{\'\i}a}, {del Pino},
  {Aparicio}, {van der Marel}  \& {Watkins}}{{Mart{\'\i}nez-Garc{\'\i}a}
  et~al.}{2021}]{Martinez-Garcia21}
{Mart{\'\i}nez-Garc{\'\i}a} A.~M.,  {del Pino} A.,  {Aparicio} A.,  {van der
  Marel} R.~P.,   {Watkins} L.~L.,  2021, \mn@doi [\mnras]
  {10.1093/mnras/stab1568}, \href
  {https://ui.adsabs.harvard.edu/abs/2021MNRAS.505.5884M} {505, 5884}

\bibitem[\protect\citeauthoryear{{Mart{\'\i}nez-V{\'a}zquez}
  et~al.,}{{Mart{\'\i}nez-V{\'a}zquez} et~al.}{2016}]{MV16}
{Mart{\'\i}nez-V{\'a}zquez} C.~E.,  et~al., 2016, \mn@doi [\mnras]
  {10.1093/mnras/stw1895}, \href
  {https://ui.adsabs.harvard.edu/abs/2016MNRAS.462.4349M} {462, 4349}

\bibitem[\protect\citeauthoryear{{Massari}, {Breddels}, {Helmi}, {Posti},
  {Brown}  \& {Tolstoy}}{{Massari} et~al.}{2018}]{Massari18}
{Massari} D.,  {Breddels} M.~A.,  {Helmi} A.,  {Posti} L.,  {Brown} A.~G.~A.,
  {Tolstoy} E.,  2018, \mn@doi [Nature Astronomy] {10.1038/s41550-017-0322-y},
  \href {https://ui.adsabs.harvard.edu/abs/2018NatAs...2..156M} {2, 156}

\bibitem[\protect\citeauthoryear{{Massari}, {Helmi}, {Mucciarelli}, {Sales},
  {Spina}  \& {Tolstoy}}{{Massari} et~al.}{2020}]{Massari20}
{Massari} D.,  {Helmi} A.,  {Mucciarelli} A.,  {Sales} L.~V.,  {Spina} L.,
  {Tolstoy} E.,  2020, \mn@doi [\aap] {10.1051/0004-6361/201935613}, \href
  {https://ui.adsabs.harvard.edu/abs/2020A&A...633A..36M} {633, A36}

\bibitem[\protect\citeauthoryear{{Mateo}}{{Mateo}}{1998}]{Mateo:1998}
{Mateo} M.~L.,  1998, \mn@doi [\araa] {10.1146/annurev.astro.36.1.435}, \href
  {https://ui.adsabs.harvard.edu/abs/1998ARA&A..36..435M} {36, 435}

\bibitem[\protect\citeauthoryear{{McConnachie}}{{McConnachie}}{2012}]{McConnachie12}
{McConnachie} A.~W.,  2012, \mn@doi [\aj] {10.1088/0004-6256/144/1/4}, \href
  {https://ui.adsabs.harvard.edu/abs/2012AJ....144....4M} {144, 4}

\bibitem[\protect\citeauthoryear{{McConnachie} \& {Venn}}{{McConnachie} \&
  {Venn}}{2020a}]{McConnachie_Venn20b}
{McConnachie} A.~W.,  {Venn} K.~A.,  2020a, \mn@doi [Research Notes of the
  American Astronomical Society] {10.3847/2515-5172/abd18b}, \href
  {https://ui.adsabs.harvard.edu/abs/2020RNAAS...4..229M} {4, 229}

\bibitem[\protect\citeauthoryear{{McConnachie} \& {Venn}}{{McConnachie} \&
  {Venn}}{2020b}]{McConnachie_Venn20}
{McConnachie} A.~W.,  {Venn} K.~A.,  2020b, \mn@doi [\aj]
  {10.3847/1538-3881/aba4ab}, \href
  {https://ui.adsabs.harvard.edu/abs/2020AJ....160..124M} {160, 124}

\bibitem[\protect\citeauthoryear{{Medina} et~al.,}{{Medina}
  et~al.}{2018}]{Medina18}
{Medina} G.~E.,  et~al., 2018, \mn@doi [\apj] {10.3847/1538-4357/aaad02}, \href
  {https://ui.adsabs.harvard.edu/abs/2018ApJ...855...43M} {855, 43}

\bibitem[\protect\citeauthoryear{{Miyamoto} \& {Nagai}}{{Miyamoto} \&
  {Nagai}}{1975}]{Miyamoto:1975}
{Miyamoto} M.,  {Nagai} R.,  1975, \pasj, \href
  {https://ui.adsabs.harvard.edu/abs/1975PASJ...27..533M} {27, 533}

\bibitem[\protect\citeauthoryear{{Montuori}, {Capuzzo-Dolcetta}, {Di Matteo},
  {Lepinette}  \& {Miocchi}}{{Montuori} et~al.}{2007}]{Montuori:2007}
{Montuori} M.,  {Capuzzo-Dolcetta} R.,  {Di Matteo} P.,  {Lepinette} A.,
  {Miocchi} P.,  2007, \mn@doi [\apj] {10.1086/512114}, \href
  {https://ui.adsabs.harvard.edu/abs/2007ApJ...659.1212M} {659, 1212}

\bibitem[\protect\citeauthoryear{{Mu{\~n}oz} et~al.,}{{Mu{\~n}oz}
  et~al.}{2005}]{2005ApJ...631L.137M}
{Mu{\~n}oz} R.~R.,  et~al., 2005, \mn@doi [\apjl] {10.1086/497396}, \href
  {https://ui.adsabs.harvard.edu/abs/2005ApJ...631L.137M} {631, L137}

\bibitem[\protect\citeauthoryear{{Mu{\~n}oz} et~al.,}{{Mu{\~n}oz}
  et~al.}{2006a}]{2006ApJ...649..201M}
{Mu{\~n}oz} R.~R.,  et~al., 2006a, \mn@doi [\apj] {10.1086/505620}, \href
  {https://ui.adsabs.harvard.edu/abs/2006ApJ...649..201M} {649, 201}

\bibitem[\protect\citeauthoryear{{Mu{\~n}oz} et~al.,}{{Mu{\~n}oz}
  et~al.}{2006b}]{Munoz06}
{Mu{\~n}oz} R.~R.,  et~al., 2006b, \mn@doi [\apj] {10.1086/505620}, \href
  {https://ui.adsabs.harvard.edu/abs/2006ApJ...649..201M} {649, 201}

\bibitem[\protect\citeauthoryear{{Mu{\~n}oz}, {C{\^o}t{\'e}}, {Santana},
  {Geha}, {Simon}, {Oyarz{\'u}n}, {Stetson}  \& {Djorgovski}}{{Mu{\~n}oz}
  et~al.}{2018}]{Munoz18}
{Mu{\~n}oz} R.~R.,  {C{\^o}t{\'e}} P.,  {Santana} F.~A.,  {Geha} M.,  {Simon}
  J.~D.,  {Oyarz{\'u}n} G.~A.,  {Stetson} P.~B.,   {Djorgovski} S.~G.,  2018,
  \mn@doi [\apj] {10.3847/1538-4357/aac16b}, \href
  {https://ui.adsabs.harvard.edu/abs/2018ApJ...860...66M} {860, 66}

\bibitem[\protect\citeauthoryear{Munoz, Majewski  \& Johnston}{Munoz
  et~al.}{2008}]{Munoz:2007jn}
Munoz R.~R.,  Majewski S.~R.,   Johnston K.~V.,  2008, \mn@doi [Astrophys. J.]
  {10.1086/587125}, 679, 346

\bibitem[\protect\citeauthoryear{{Muraveva}, {Clementini}, {Garofalo}  \&
  {Cusano}}{{Muraveva} et~al.}{2020}]{Muraveva20}
{Muraveva} T.,  {Clementini} G.,  {Garofalo} A.,   {Cusano} F.,  2020, \mn@doi
  [\mnras] {10.1093/mnras/staa2984}, \href
  {https://ui.adsabs.harvard.edu/abs/2020MNRAS.499.4040M} {499, 4040}

\bibitem[\protect\citeauthoryear{{Mutlu-Pakdil} et~al.,}{{Mutlu-Pakdil}
  et~al.}{2018}]{Mutlu-Pakdil18}
{Mutlu-Pakdil} B.,  et~al., 2018, \mn@doi [\apj] {10.3847/1538-4357/aacd0e},
  \href {https://ui.adsabs.harvard.edu/abs/2018ApJ...863...25M} {863, 25}

\bibitem[\protect\citeauthoryear{{Mutlu-Pakdil} et~al.,}{{Mutlu-Pakdil}
  et~al.}{2019}]{Mutlu-Pakdil19}
{Mutlu-Pakdil} B.,  et~al., 2019, \mn@doi [\apj] {10.3847/1538-4357/ab45ec},
  \href {https://ui.adsabs.harvard.edu/abs/2019ApJ...885...53M} {885, 53}

\bibitem[\protect\citeauthoryear{{Mutlu-Pakdil} et~al.,}{{Mutlu-Pakdil}
  et~al.}{2020}]{Mutlu-Pakdil20}
{Mutlu-Pakdil} B.,  et~al., 2020, \mn@doi [\apj] {10.3847/1538-4357/abb40b},
  \href {https://ui.adsabs.harvard.edu/abs/2020ApJ...902..106M} {902, 106}

\bibitem[\protect\citeauthoryear{{Navarro}, {Frenk}  \& {White}}{{Navarro}
  et~al.}{1996}]{Navarro:1996}
{Navarro} J.~F.,  {Frenk} C.~S.,   {White} S. D.~M.,  1996, \mn@doi [\apj]
  {10.1086/177173}, \href
  {https://ui.adsabs.harvard.edu/abs/1996ApJ...462..563N} {462, 563}

\bibitem[\protect\citeauthoryear{{Newberg} et~al.,}{{Newberg}
  et~al.}{2002}]{Newberg02}
{Newberg} H.~J.,  et~al., 2002, \mn@doi [\apj] {10.1086/338983}, \href
  {https://ui.adsabs.harvard.edu/abs/2002ApJ...569..245N} {569, 245}

\bibitem[\protect\citeauthoryear{{Pace} \& {Li}}{{Pace} \&
  {Li}}{2019}]{Pace&Li19}
{Pace} A.~B.,  {Li} T.~S.,  2019, \mn@doi [\apj] {10.3847/1538-4357/ab0aee},
  \href {https://ui.adsabs.harvard.edu/abs/2019ApJ...875...77P} {875, 77}

\bibitem[\protect\citeauthoryear{{Pace} et~al.,}{{Pace} et~al.}{2020}]{Pace20}
{Pace} A.~B.,  et~al., 2020, \mn@doi [\mnras] {10.1093/mnras/staa1419}, \href
  {https://ui.adsabs.harvard.edu/abs/2020MNRAS.495.3022P} {495, 3022}

\bibitem[\protect\citeauthoryear{{Pace}, {Walker}, {Koposov}, {Caldwell},
  {Mateo}, {Olszewski}, {Bailey}  \& {Wang}}{{Pace} et~al.}{2021}]{Pace21}
{Pace} A.~B.,  {Walker} M.~G.,  {Koposov} S.~E.,  {Caldwell} N.,  {Mateo} M.,
  {Olszewski} E.~W.,  {Bailey} John~I. I.,   {Wang} M.-Y.,  2021, arXiv
  e-prints, \href {https://ui.adsabs.harvard.edu/abs/2021arXiv210500064P} {p.
  arXiv:2105.00064}

\bibitem[\protect\citeauthoryear{{Palaversa} et~al.,}{{Palaversa}
  et~al.}{2013}]{Palaversa13}
{Palaversa} L.,  et~al., 2013, \mn@doi [\aj] {10.1088/0004-6256/146/4/101},
  \href {https://ui.adsabs.harvard.edu/abs/2013AJ....146..101P} {146, 101}

\bibitem[\protect\citeauthoryear{Penarrubia, Navarro, McConnachie  \&
  Martin}{Penarrubia et~al.}{2009}]{Penarrubia:2008mu}
Penarrubia J.,  Navarro J.~F.,  McConnachie A.~W.,   Martin N.~F.,  2009,
  \mn@doi [Astrophys. J.] {10.1088/0004-637X/698/1/222}, 698, 222

\bibitem[\protect\citeauthoryear{{Plummer}}{{Plummer}}{1911}]{Plummer1911}
{Plummer} H.~C.,  1911, \mn@doi [\mnras] {10.1093/mnras/71.5.460}, \href
  {https://ui.adsabs.harvard.edu/abs/1911MNRAS..71..460P} {71, 460}

\bibitem[\protect\citeauthoryear{Price-Whelan}{Price-Whelan}{2017}]{gala}
Price-Whelan A.~M.,  2017, \mn@doi [The Journal of Open Source Software]
  {10.21105/joss.00388}, 2

\bibitem[\protect\citeauthoryear{{Price-Whelan} et~al.,}{{Price-Whelan}
  et~al.}{2020}]{gala-v1.3}
{Price-Whelan} A.,  et~al., 2020, {adrn/gala: v1.3},
  \mn@doi{10.5281/zenodo.4159870}

\bibitem[\protect\citeauthoryear{{Reid} \& {Brunthaler}}{{Reid} \&
  {Brunthaler}}{2004}]{Reid:2004}
{Reid} M.~J.,  {Brunthaler} A.,  2004, \mn@doi [\apj] {10.1086/424960}, \href
  {https://ui.adsabs.harvard.edu/abs/2004ApJ...616..872R} {616, 872}

\bibitem[\protect\citeauthoryear{{Riello} et~al.,}{{Riello}
  et~al.}{2021}]{Riello21}
{Riello} M.,  et~al., 2021, \mn@doi [\aap] {10.1051/0004-6361/202039587}, \href
  {https://ui.adsabs.harvard.edu/abs/2021A&A...649A...3R} {649, A3}

\bibitem[\protect\citeauthoryear{Rocha, Peter  \& Bullock}{Rocha
  et~al.}{2012}]{Rocha:2011aa}
Rocha M.,  Peter A. H.~G.,   Bullock J.~S.,  2012, \mn@doi [Mon. Not. Roy.
  Astron. Soc.] {10.1111/j.1365-2966.2012.21432.x}, 425, 231

\bibitem[\protect\citeauthoryear{{Samus'}, {Kazarovets}, {Durlevich}, {Kireeva}
   \& {Pastukhova}}{{Samus'} et~al.}{2017}]{Samus17}
{Samus'} N.~N.,  {Kazarovets} E.~V.,  {Durlevich} O.~V.,  {Kireeva} N.~N.,
  {Pastukhova} E.~N.,  2017, \mn@doi [Astronomy Reports]
  {10.1134/S1063772917010085}, \href
  {https://ui.adsabs.harvard.edu/abs/2017ARep...61...80S} {61, 80}

\bibitem[\protect\citeauthoryear{{Sesar} et~al.,}{{Sesar}
  et~al.}{2017}]{Sesar17}
{Sesar} B.,  et~al., 2017, \mn@doi [\aj] {10.3847/1538-3881/aa661b}, \href
  {https://ui.adsabs.harvard.edu/abs/2017AJ....153..204S} {153, 204}

\bibitem[\protect\citeauthoryear{{Shipp} et~al.,}{{Shipp}
  et~al.}{2018}]{Shipp18}
{Shipp} N.,  et~al., 2018, \mn@doi [\apj] {10.3847/1538-4357/aacdab}, \href
  {https://ui.adsabs.harvard.edu/abs/2018ApJ...862..114S} {862, 114}

\bibitem[\protect\citeauthoryear{{Simon}}{{Simon}}{2018}]{Simon18}
{Simon} J.~D.,  2018, \mn@doi [\apj] {10.3847/1538-4357/aacdfb}, \href
  {https://ui.adsabs.harvard.edu/abs/2018ApJ...863...89S} {863, 89}

\bibitem[\protect\citeauthoryear{{Simon}}{{Simon}}{2019}]{Simon19}
{Simon} J.~D.,  2019, \mn@doi [\araa] {10.1146/annurev-astro-091918-104453},
  \href {https://ui.adsabs.harvard.edu/abs/2019ARA&A..57..375S} {57, 375}

\bibitem[\protect\citeauthoryear{{Spencer}, {Mateo}, {Olszewski}, {Walker},
  {McConnachie}  \& {Kirby}}{{Spencer} et~al.}{2018}]{Spencer18}
{Spencer} M.~E.,  {Mateo} M.,  {Olszewski} E.~W.,  {Walker} M.~G.,
  {McConnachie} A.~W.,   {Kirby} E.~N.,  2018, \mn@doi [\aj]
  {10.3847/1538-3881/aae3e4}, \href
  {https://ui.adsabs.harvard.edu/abs/2018AJ....156..257S} {156, 257}

\bibitem[\protect\citeauthoryear{Strigari, Frenk  \& White}{Strigari
  et~al.}{2018}]{Strigari:2018bcn}
Strigari L.~E.,  Frenk C.~S.,   White S. D.~M.,  2018, \mn@doi [Astrophys. J.]
  {10.3847/1538-4357/aac2d3}, 860, 56

\bibitem[\protect\citeauthoryear{{Stringer} et~al.,}{{Stringer}
  et~al.}{2021}]{Stringer21}
{Stringer} K.~M.,  et~al., 2021, \mn@doi [\apj] {10.3847/1538-4357/abe873},
  \href {https://ui.adsabs.harvard.edu/abs/2021ApJ...911..109S} {911, 109}

\bibitem[\protect\citeauthoryear{{Torrealba} et~al.,}{{Torrealba}
  et~al.}{2015}]{Torrealba15}
{Torrealba} G.,  et~al., 2015, \mn@doi [\mnras] {10.1093/mnras/stu2274}, \href
  {https://ui.adsabs.harvard.edu/abs/2015MNRAS.446.2251T} {446, 2251}

\bibitem[\protect\citeauthoryear{{Ural}, {Wilkinson}, {Read}  \&
  {Walker}}{{Ural} et~al.}{2015}]{2015NatCo...6.7599U}
{Ural} U.,  {Wilkinson} M.~I.,  {Read} J.~I.,   {Walker} M.~G.,  2015, \mn@doi
  [Nature Communications] {10.1038/ncomms8599}, \href
  {https://ui.adsabs.harvard.edu/abs/2015NatCo...6.7599U} {6, 7599}

\bibitem[\protect\citeauthoryear{{Vasiliev}}{{Vasiliev}}{2019}]{Vasiliev19b}
{Vasiliev} E.,  2019, \mn@doi [\mnras] {10.1093/mnras/stz2100}, \href
  {https://ui.adsabs.harvard.edu/abs/2019MNRAS.489..623V} {489, 623}

\bibitem[\protect\citeauthoryear{Virtanen et~al.,}{Virtanen
  et~al.}{2020}]{SciPy20}
Virtanen P.,  et~al., 2020, \mn@doi [Nature Methods]
  {10.1038/s41592-019-0686-2}, \href {https://rdcu.be/b08Wh} {17, 261}

\bibitem[\protect\citeauthoryear{{Vivas} \& {Mateo}}{{Vivas} \&
  {Mateo}}{2013}]{Vivas13}
{Vivas} A.~K.,  {Mateo} M.,  2013, \mn@doi [\aj] {10.1088/0004-6256/146/6/141},
  \href {https://ui.adsabs.harvard.edu/abs/2013AJ....146..141V} {146, 141}

\bibitem[\protect\citeauthoryear{{Vivas} et~al.,}{{Vivas}
  et~al.}{2004}]{Vivas04}
{Vivas} A.~K.,  et~al., 2004, \mn@doi [\aj] {10.1086/380929}, \href
  {https://ui.adsabs.harvard.edu/abs/2004AJ....127.1158V} {127, 1158}

\bibitem[\protect\citeauthoryear{{Vivas}, {Alonso-Garc{\'\i}a}, {Mateo},
  {Walker}  \& {Howard}}{{Vivas} et~al.}{2019}]{Vivas19}
{Vivas} A.~K.,  {Alonso-Garc{\'\i}a} J.,  {Mateo} M.,  {Walker} A.,   {Howard}
  B.,  2019, \mn@doi [\aj] {10.3847/1538-3881/aaf4f3}, \href
  {https://ui.adsabs.harvard.edu/abs/2019AJ....157...35V} {157, 35}

\bibitem[\protect\citeauthoryear{{Walker}, {Mateo}, {Olszewski}, {Gnedin},
  {Wang}, {Sen}  \& {Woodroofe}}{{Walker} et~al.}{2007}]{2007ApJ...667L..53W}
{Walker} M.~G.,  {Mateo} M.,  {Olszewski} E.~W.,  {Gnedin} O.~Y.,  {Wang} X.,
  {Sen} B.,   {Woodroofe} M.,  2007, \mn@doi [\apjl] {10.1086/521998}, \href
  {https://ui.adsabs.harvard.edu/abs/2007ApJ...667L..53W} {667, L53}

\bibitem[\protect\citeauthoryear{{Walker}, {Mateo}  \& {Olszewski}}{{Walker}
  et~al.}{2009}]{Walker09}
{Walker} M.~G.,  {Mateo} M.,   {Olszewski} E.~W.,  2009, \mn@doi [\aj]
  {10.1088/0004-6256/137/2/3100}, \href
  {https://ui.adsabs.harvard.edu/abs/2009AJ....137.3100W} {137, 3100}

\bibitem[\protect\citeauthoryear{{Walker}, {Olszewski}  \& {Mateo}}{{Walker}
  et~al.}{2015}]{Walker15}
{Walker} M.~G.,  {Olszewski} E.~W.,   {Mateo} M.,  2015, \mn@doi [\mnras]
  {10.1093/mnras/stv099}, \href
  {https://ui.adsabs.harvard.edu/abs/2015MNRAS.448.2717W} {448, 2717}

\bibitem[\protect\citeauthoryear{{Wang} et~al.,}{{Wang} et~al.}{2017}]{Wang17}
{Wang} M.~Y.,  et~al., 2017, \mn@doi [\mnras] {10.1093/mnras/stx742}, \href
  {https://ui.adsabs.harvard.edu/abs/2017MNRAS.468.4887W} {468, 4887}

\bibitem[\protect\citeauthoryear{{Wang} et~al.,}{{Wang}
  et~al.}{2019}]{2019ApJ...881..118W}
{Wang} M.~Y.,  et~al., 2019, \mn@doi [\apj] {10.3847/1538-4357/ab31a9}, \href
  {https://ui.adsabs.harvard.edu/abs/2019ApJ...881..118W} {881, 118}

\makeatother
\end{thebibliography}

\section*{Appendix}

This appendix provides information on the properties of the bright ($G < 19$) member candidates beyond the King limiting radius in each of the dSphs that we study in Tables~\ref{tab:Member} to \ref{tab:Member_C3}.

\tabcolsep=0.15cm
\begin{table*}
\centering
\caption{Possible member candidates beyond the King limiting radius for Draco and Ursa Minor. We only include stars with $P_{PM}\geq0.9$, and the dust-corrected G magnitude: $G\leq19$ mag. This table will be made available in its entirety in machine-readable form. The values for radial velocity $V_{r}$ are taken from the work cited in Section \ref{sec:RV}.}  

\label{tab:Member}	
\begin{tabular}{lcccccccrr} 
\hline
Galaxy & Source ID & $\alpha$ & $\delta$ &  $P_{Space+PM}$ &  $P_{PM}$ & $V_{r}$ &   $G$ &  $\mu_{\alpha}cos\delta$ &     $\mu_{\delta}$ \\
 
 & & [deg] & [deg] & & & [${\rm km~s^{-1}}$] &   [mag] & [${\rm mas~yr^{-1}}$] & [${\rm mas~yr^{-1}}$] \\
\hline
Draco       &  1434492516786370176 &  261.909753 &  58.263091 &        0.864 &       0.978 &  $-288.2$    &  17.77 &   $0.047\pm0.127$ &  $-0.217\pm0.138$ \\
      &  1433949770359264768 &  259.142130 &  58.498825 &        0.863 &       0.958 &   $-292.3$     &  18.68 &   $0.093\pm0.187$ &  $-0.206\pm0.172$ \\
      &  1434014607185724544 &  260.617774 &  58.730676 &        0.600 &       0.936 &    -      &  18.75 &  $-0.084\pm0.186$ &  $-0.152\pm0.197$ \\
      &  1433509931348820864 &  256.354218 &  57.552936 &        0.129 &       0.910 &       -     &  18.96 &   $0.205\pm0.184$ &  $-0.259\pm0.281$ \\
      &  1422046732355091584 &  262.992811 &  56.621840 &        0.087 &       0.938 &      -      &  18.35 &  $-0.093\pm0.141$ &  $-0.311\pm0.147$ \\
      &  1420524699025187328 &  261.600665 &  56.016370 &        0.048 &       0.937 &      -      &  17.89 &  $-0.134\pm0.116$ &  $-0.205\pm0.110$ \\
      &  1435623055258755840 &  262.155953 &  59.890716 &        0.042 &       0.943 &       -     &  18.43 &  $-0.057\pm0.168$ &  $-0.066\pm0.172$ \\
      &  1437346161777223936 &  258.784927 &  60.251350 &        0.022 &       0.931 &       -     &  18.99 &  $-0.071\pm0.202$ &  $-0.152\pm0.227$ \\
      &  1432365034801179648 &  256.952365 &  55.751991 &        0.014 &       0.909 &       -     &  18.69 &  $-0.149\pm0.182$ &  $-0.159\pm0.229$ \\
      &  1437470136009868416 &  260.717683 &  60.847082 &        0.009 &       0.928 &      -      &  18.96 &  $-0.009\pm0.221$ &  $-0.169\pm0.251$ \\

\hline
Ursa Minor       &  1645171112311713536 &  228.351877 &  66.422532 &        0.990 &       0.999 &        - &  16.14 &  $-0.145\pm0.041$ &   $0.066\pm0.044$ \\
      &  1645606828153598848 &  230.260134 &  67.470024 &        0.988 &       0.997 &        - &  16.77 &  $-0.162\pm0.059$ &   $0.069\pm0.064$ \\
      &  1693470798398323328 &  225.704777 &  67.572588 &        0.988 &       0.995 &        - &  17.12 &  $-0.214\pm0.068$ &   $0.099\pm0.065$ \\
      &  1645270510739088000 &  227.326351 &  66.442343 &        0.985 &       0.994 &      $-256.8$ &  17.85 &  $-0.097\pm0.092$ &  $-0.012\pm0.097$ \\
      &  1645194511293583616 &  229.300906 &  66.777715 &        0.983 &       0.997 &        - &  16.42 &  $-0.162\pm0.046$ &   $0.042\pm0.048$ \\
      &  1645185092429838592 &  228.491460 &  66.710679 &        0.976 &       0.991 &        - &  17.19 &  $-0.183\pm0.077$ &   $0.177\pm0.080$ \\
      &  1645204196444361728 &  228.979185 &  66.690191 &        0.974 &       0.995 &        - &  17.25 &  $-0.115\pm0.080$ &   $0.002\pm0.078$ \\
      &  1645635995276558336 &  230.874290 &  67.670574 &        0.972 &       0.995 &        - &  17.12 &  $-0.168\pm0.059$ &   $0.136\pm0.069$ \\
      &  1693464785444020224 &  224.677313 &  67.359829 &        0.936 &       0.988 &        - &  17.78 &  $-0.173\pm0.094$ &  $-0.061\pm0.105$ \\
      &  1693573430936780032 &  226.089830 &  67.779650 &        0.926 &       0.973 &        - &  17.88 &  $-0.220\pm0.125$ &   $0.252\pm0.107$ \\
      &  1693459013010166400 &  224.989988 &  67.202354 &        0.926 &       0.961 &        - &  18.91 &   $0.137\pm0.216$ &   $0.050\pm0.201$ \\
      &  1645948119139534336 &  230.439490 &  68.295812 &        0.919 &       0.985 &        - &  18.24 &  $-0.217\pm0.156$ &   $0.058\pm0.137$ \\
      &  1669324938936435200 &  224.507557 &  66.213615 &        0.919 &       0.978 &        - &  18.21 &  $-0.293\pm0.142$ &   $0.168\pm0.135$ \\
      &  1693484061257345792 &  224.702610 &  67.651446 &        0.866 &       0.986 &        - &  17.95 &  $-0.267\pm0.102$ &   $0.136\pm0.102$ \\
      &  1645527727740223744 &  230.409902 &  66.895595 &        0.850 &       0.987 &        - &  18.58 &  $-0.095\pm0.156$ &   $0.151\pm0.146$ \\
      &  1645257110440749696 &  227.204801 &  66.152029 &        0.836 &       0.973 &        - &  18.94 &  $-0.077\pm0.193$ &  $-0.095\pm0.209$ \\
      &  1645678017235803392 &  232.362738 &  67.876246 &        0.822 &       0.989 &        - &  18.33 &  $-0.075\pm0.123$ &   $0.196\pm0.133$ \\
      &  1693681908926278016 &  224.945984 &  67.943088 &        0.731 &       0.979 &        - &  18.81 &  $-0.193\pm0.196$ &   $0.170\pm0.193$ \\
      &  1693457849072756608 &  224.336427 &  67.494067 &        0.704 &       0.967 &        - &  18.90 &  $-0.266\pm0.188$ &  $-0.080\pm0.193$ \\
      &  1693636382271997056 &  223.382839 &  67.272458 &        0.698 &       0.977 &        - &  18.69 &  $-0.148\pm0.166$ &   $0.254\pm0.180$ \\
      &  1693671601004793984 &  224.122415 &  67.667227 &        0.692 &       0.977 &        - &  18.61 &  $-0.158\pm0.187$ &   $0.219\pm0.200$ \\
      &  1645155062017982848 &  228.595163 &  66.300722 &        0.636 &       0.962 &        - &  18.94 &  $-0.216\pm0.208$ &  $-0.109\pm0.232$ \\
      &  1669596518308786816 &  222.076584 &  66.913474 &        0.599 &       0.984 &        - &  18.86 &  $-0.134\pm0.184$ &   $0.120\pm0.171$ \\
      &  1645557655075852032 &  231.167407 &  67.175327 &        0.573 &       0.960 &        - &  18.85 &  $-0.074\pm0.264$ &  $-0.106\pm0.246$ \\
      &  1645221376313145728 &  226.395026 &  65.730813 &        0.571 &       0.956 &        - &  18.61 &  $-0.298\pm0.234$ &   $0.289\pm0.191$ \\
      &  1645064764625656064 &  227.457343 &  66.023662 &        0.562 &       0.948 &        - &  18.81 &  $-0.247\pm0.200$ &  $-0.148\pm0.178$ \\
      &  1645771063407500800 &  232.008520 &  68.238955 &        0.520 &       0.947 &        - &  18.41 &  $-0.210\pm0.218$ &  $-0.107\pm0.182$ \\
      &  1645042430795689344 &  226.810358 &  65.895622 &        0.440 &       0.911 &        - &  18.40 &   $0.031\pm0.162$ &  $-0.231\pm0.145$ \\
      &  1669619608053127168 &  223.195017 &  67.235376 &        0.336 &       0.930 &        - &  18.66 &  $-0.368\pm0.160$ &  $-0.094\pm0.192$ \\
      &  1621087886357153920 &  223.245645 &  64.770400 &        0.096 &       0.908 &        - &  18.68 &  $-0.428\pm0.167$ &  $-0.094\pm0.157$ \\

\hline
\end{tabular}
\end{table*}

\tabcolsep=0.15cm
\begin{table*}
\centering
\caption{Continued: possible bright ($G \leq 19$ mag) member candidates beyond the King limiting radius for Sextans and Sculptor.}   

\label{tab:Member_C1}	
\begin{tabular}{lcccccccrr} 
\hline
Galaxy & Source ID & $\alpha$ & $\delta$ &  $P_{Space+PM}$ &  $P_{PM}$ & $V_{r}$ &   $G$ &  $\mu_{\alpha}cos\delta$ &     $\mu_{\delta}$ \\
 
 & & [deg] & [deg] & & & [${\rm km~s^{-1}}$] &   [mag] & [${\rm mas~yr^{-1}}$] & [${\rm mas~yr^{-1}}$] \\
\hline
Sextans       &  3828856623037326464 &  153.378402 & -2.388383 &        0.976 &       0.966 &        - &  18.04 &  $-0.308\pm0.138$ &   $0.192\pm0.143$ \\
      &  3829914043985259904 &  152.454533 & -1.442343 &        0.975 &       0.970 &        - &  18.04 &  $-0.358\pm0.171$ &  $-0.118\pm0.147$ \\
      &  3830628451665548928 &  154.323570 & -1.460667 &        0.972 &       0.979 &        - &  17.77 &  $-0.513\pm0.162$ &   $0.177\pm0.223$ \\
      &  3830708647294573824 &  154.132163 & -0.962963 &        0.972 &       0.976 &        - &  17.28 &  $-0.561\pm0.120$ &   $0.021\pm0.133$ \\
      &  3828753990498353792 &  153.066253 & -2.583271 &        0.969 &       0.976 &        - &  18.58 &  $-0.365\pm0.192$ &  $-0.008\pm0.182$ \\
      &  3830604331128780672 &  154.232532 & -1.519406 &        0.954 &       0.958 &        - &  17.58 &  $-0.215\pm0.149$ &  $-0.159\pm0.209$ \\
      &  3831478820830542592 &  153.555006 & -0.703581 &        0.953 &       0.949 &        - &  18.85 &  $-0.267\pm0.260$ &   $0.191\pm0.229$ \\
      &  3830319390113933952 &  154.654507 & -2.160077 &        0.950 &       0.993 &        - &  16.76 &  $-0.411\pm0.088$ &   $0.106\pm0.085$ \\
      &  3830545339753349120 &  154.821339 & -1.192598 &        0.948 &       0.990 &        - &  17.50 &  $-0.406\pm0.109$ &   $0.109\pm0.109$ \\
      &  3828858959499558144 &  153.494908 & -2.299684 &        0.948 &       0.903 &        - &  18.91 &  $-0.609\pm0.326$ &   $0.585\pm0.462$ \\
      &  3830401608672381440 &  154.167771 & -1.802013 &        0.947 &       0.953 &        - &  18.91 &  $-0.257\pm0.271$ &   $0.009\pm0.251$ \\
      &  3828950566856844288 &  153.172068 & -2.426458 &        0.945 &       0.915 &        - &  17.88 &  $-0.314\pm0.132$ &   $0.268\pm0.124$ \\
      &  3830390720930784640 &  154.500820 & -1.922632 &        0.943 &       0.985 &        - &  16.88 &  $-0.306\pm0.104$ &  $-0.062\pm0.152$ \\
      &  3829950705825803136 &  152.749148 & -1.067950 &        0.922 &       0.915 &        231.53 &  18.67 &  $-0.630\pm0.213$ &  $-0.113\pm0.207$ \\
      &  3828784987277714560 &  153.925853 & -2.644071 &        0.913 &       0.980 &        - &  17.01 &  $-0.544\pm0.090$ &   $0.154\pm0.095$ \\
      &  3830602338263932928 &  154.212478 & -1.583762 &        0.910 &       0.918 &        - &  18.27 &  $-0.602\pm0.226$ &  $-0.077\pm0.262$ \\
      &  3830816949190286080 &  154.618951 & -0.547751 &        0.902 &       0.982 &        226.98 &  18.13 &  $-0.440\pm0.179$ &   $0.059\pm0.147$ \\
      &  3831492186768793600 &  153.438987 & -0.626571 &        0.880 &       0.923 &        231.39 &  18.54 &  $-0.591\pm0.226$ &  $-0.147\pm0.211$ \\
      &  3830445181116015872 &  154.681072 & -1.391989 &        0.877 &       0.963 &        - &  17.64 &  $-0.564\pm0.146$ &   $0.263\pm0.157$ \\
      &  3831551732195432576 &  153.463839 & -0.440007 &        0.877 &       0.944 &        237.0 &  18.65 &  $-0.522\pm0.263$ &   $0.277\pm0.249$ \\
      &  3828710083048013952 &  153.109300 & -2.888955 &        0.873 &       0.959 &        - &  18.65 &  $-0.241\pm0.240$ &   $0.135\pm0.205$ \\
      &  3830740292613833856 &  154.685360 & -1.027934 &        0.870 &       0.965 &        - &  18.44 &  $-0.403\pm0.246$ &   $0.103\pm0.229$ \\
      &  3831535553053384704 &  152.872744 & -0.523191 &        0.861 &       0.959 &        - &  18.19 &  $-0.208\pm0.199$ &  $-0.133\pm0.204$ \\
      &  3831513459741694848 &  152.810584 & -0.740800 &        0.829 &       0.910 &        - &  18.68 &  $-0.181\pm0.355$ &  $-0.066\pm0.434$ \\
      &  3830337287242365440 &  154.469507 & -1.993393 &        0.754 &       0.927 &        - &  18.50 &  $-0.160\pm0.261$ &  $-0.149\pm0.347$ \\
      &  3831759406748927744 &  152.885667 & -0.148648 &        0.753 &       0.967 &        - &  18.24 &  $-0.222\pm0.184$ &  $-0.065\pm0.175$ \\
      &  3831812247731524608 &  152.006430 &  0.018912 &        0.662 &       0.985 &        - &  17.78 &  $-0.410\pm0.139$ &   $0.121\pm0.123$ \\
      &  3780600603882398592 &  153.920001 & -3.463134 &        0.543 &       0.968 &        - &  18.04 &  $-0.469\pm0.231$ &   $0.207\pm0.281$ \\
      &  3828582745857545856 &  151.638894 & -2.787752 &        0.446 &       0.908 &        - &  17.96 &  $-0.212\pm0.231$ &  $-0.311\pm0.262$ \\
      &  3831259678714011008 &  155.039181 &  0.201047 &        0.278 &       0.922 &        - &  18.10 &  $-0.233\pm0.174$ &  $-0.265\pm0.172$ \\
      &  3833327202955907584 &  151.378337 &  0.067266 &        0.226 &       0.956 &        - &  18.02 &  $-0.173\pm0.160$ &  $-0.019\pm0.186$ \\
      &  3828376926729977984 &  151.580187 & -3.824589 &        0.214 &       0.930 &        - &  18.40 &  $-0.398\pm0.216$ &  $-0.196\pm0.237$ \\
      &  3828173379639671680 &  151.843216 & -3.932880 &        0.174 &       0.906 &        - &  18.74 &  $-0.198\pm0.336$ &  $-0.140\pm0.468$ \\
      &  3832339394837569152 &  154.075660 &  1.169582 &        0.127 &       0.913 &        - &  18.70 &  $-0.745\pm0.239$ &   $0.149\pm0.246$ \\

\hline
Sculptor       &  5026328277217292416 &  16.299378 & -34.384139 &        0.992 &       0.999 &        95.78 &  18.27 &   $0.186\pm0.119$ &  $-0.098\pm0.103$ \\
      &  5027479362811922048 &  14.921115 & -32.834462 &        0.987 &       0.997 &        - &  18.27 &  $-0.142\pm0.135$ &  $-0.087\pm0.090$ \\
      &  5026130884816022016 &  16.843514 & -34.664602 &        0.985 &       1.000 &        118.78 &  17.82 &   $0.039\pm0.089$ &  $-0.088\pm0.076$ \\
      &  5002282889926690944 &  15.894926 & -34.914193 &        0.980 &       0.999 &        - &  17.40 &   $0.177\pm0.082$ &  $-0.241\pm0.066$ \\
      &  5006419626331394048 &  12.675394 & -33.468379 &        0.980 &       0.999 &        - &  17.59 &   $0.077\pm0.103$ &  $-0.176\pm0.121$ \\
      &  5026553230423938816 &  16.880088 & -33.216605 &        0.964 &       0.998 &        - &  18.78 &  $-0.093\pm0.147$ &  $-0.147\pm0.139$ \\
      &  5003074675736263936 &  13.984518 & -34.337675 &        0.930 &       0.991 &        - &  18.78 &   $0.313\pm0.200$ &  $-0.459\pm0.185$ \\
      &  5006627262230333568 &  11.912982 & -32.941090 &        0.660 &       0.995 &        - &  18.94 &   $0.008\pm0.190$ &   $0.138\pm0.235$ \\
      &  5002261342074627968 &  15.575382 & -35.183082 &        0.514 &       0.980 &        - &  18.78 &   $0.365\pm0.105$ &  $-0.018\pm0.136$ \\
      &  4989939561938365056 &  16.437265 & -36.393187 &        0.021 &       0.920 &        - &  18.92 &   $0.457\pm0.127$ &   $0.030\pm0.156$ \\

\hline
\end{tabular}
\end{table*}

\tabcolsep=0.15cm
\begin{table*}
\centering
\caption{Continued: possible bright ($G \leq 19$ mag) member candidates beyond the King limiting radius for Fornax.}  
\label{tab:Member_C2}	
\begin{tabular}{lcccccccrr} 
\hline
Galaxy & Source ID & $\alpha$ & $\delta$ &  $P_{Space+PM}$ &  $P_{PM}$ & $V_{r}$ &   $G$ &  $\mu_{\alpha}cos\delta$ &     $\mu_{\delta}$ \\
 
 & & [deg] & [deg] & & & [${\rm km~s^{-1}}$] &   [mag] & [${\rm mas~yr^{-1}}$] & [${\rm mas~yr^{-1}}$] \\
\hline
Fornax      &  5050579243118343424 &  41.157190 & -34.431349 &        0.999 &       1.000 &        - &  18.24 &  $0.289\pm0.081$ &  $-0.291\pm0.117$ \\
      &  5062736616203680384 &  40.533001 & -33.205159 &        0.999 &       1.000 &        - &  18.23 &  $0.391\pm0.070$ &  $-0.313\pm0.118$ \\
      &  5062705000946431872 &  40.217350 & -33.409180 &        0.999 &       1.000 &        - &  18.42 &  $0.332\pm0.079$ &  $-0.205\pm0.137$ \\
      &  5062893812007463680 &  40.124558 & -33.341377 &        0.999 &       1.000 &        - &  18.17 &  $0.367\pm0.067$ &  $-0.480\pm0.114$ \\
      &  5050045017906217600 &  39.601695 & -35.497694 &        0.999 &       1.000 &        59.16 &  18.76 &  $0.460\pm0.123$ &  $-0.422\pm0.181$ \\
      &  5062050173052716288 &  39.425662 & -35.460194 &        0.999 &       1.000 &        70.80 &  18.64 &  $0.395\pm0.106$ &  $-0.193\pm0.160$ \\
      &  5050631126322947456 &  41.375949 & -33.930902 &        0.998 &       1.000 &        - &  18.69 &  $0.455\pm0.094$ &  $-0.259\pm0.139$ \\
      &  5062595225880065280 &  39.304869 & -33.132383 &        0.998 &       1.000 &        - &  17.87 &  $0.352\pm0.052$ &  $-0.300\pm0.095$ \\
      &  5050059998752238848 &  40.353540 & -35.314325 &        0.997 &       0.999 &       53.54 &  18.66 &  $0.289\pm0.102$ &  $-0.108\pm0.148$ \\
      &  5062921261140200704 &  39.988323 & -33.043346 &        0.997 &       1.000 &        - &  18.20 &  $0.278\pm0.064$ &  $-0.278\pm0.110$ \\
      &  5062116418628357888 &  39.009953 & -35.297152 &        0.997 &       0.999 &        - &  18.84 &  $0.374\pm0.106$ &  $-0.590\pm0.162$ \\
      &  5062400711103888000 &  38.477085 & -34.427507 &        0.997 &       1.000 &        - &  17.76 &  $0.288\pm0.054$ &  $-0.245\pm0.086$ \\
      &  5050747262238524288 &  41.914633 & -33.274775 &        0.996 &       1.000 &        - &  18.11 &  $0.409\pm0.071$ &  $-0.427\pm0.116$ \\
      &  5050146447853774592 &  41.076959 & -34.831323 &        0.996 &       0.999 &        - &  18.63 &  $0.326\pm0.100$ &  $-0.109\pm0.143$ \\
      &  5062382053765788800 &  38.886398 & -34.276573 &        0.996 &       0.999 &        - &  19.00 &  $0.276\pm0.116$ &  $-0.026\pm0.204$ \\
      &  5050008974540402304 &  40.355478 & -35.530882 &        0.995 &       0.999 &        - &  18.85 &  $0.300\pm0.129$ &  $-0.465\pm0.168$ \\
      &  5050022649716434560 &  39.921591 & -35.768959 &        0.995 &       0.999 &        - &  18.86 &  $0.236\pm0.135$ &  $-0.288\pm0.175$ \\
      &  5062334499888230400 &  38.686711 & -34.704170 &        0.995 &       0.999 &        64.02 &  17.98 &  $0.514\pm0.066$ &  $-0.489\pm0.101$ \\
      &  4965973781866201088 &  39.223780 & -35.576720 &        0.994 &       0.999 &        - &  18.98 &  $0.189\pm0.139$ &  $-0.428\pm0.193$ \\
      &  5050584981194691072 &  41.292742 & -34.361836 &        0.993 &       0.998 &        - &  18.57 &  $0.184\pm0.091$ &  $-0.193\pm0.137$ \\
      &  5063031281023014016 &  40.797560 & -32.763278 &        0.992 &       1.000 &        - &  18.50 &  $0.396\pm0.074$ &  $-0.533\pm0.132$ \\
      &  5050148200201001216 &  41.183856 & -34.796284 &        0.989 &       0.998 &        - &  18.42 &  $0.190\pm0.092$ &  $-0.307\pm0.138$ \\
      &  5062327868458705152 &  38.673459 & -34.901775 &        0.989 &       0.997 &        42.14 &  18.96 &  $0.208\pm0.125$ &   $0.062\pm0.195$ \\
      &  5050053298603169152 &  40.058157 & -35.496765 &        0.983 &       0.996 &        - &  18.65 &  $0.558\pm0.112$ &  $-0.663\pm0.150$ \\
      &  5049893113502647552 &  41.017739 & -35.709940 &        0.981 &       0.999 &        - &  18.91 &  $0.257\pm0.137$ &  $-0.383\pm0.182$ \\
      &  5049899126456915584 &  41.167672 & -35.628223 &        0.976 &       0.999 &        - &  18.83 &  $0.436\pm0.130$ &  $-0.531\pm0.181$ \\
      &  4966282744633038976 &  37.596957 & -34.988158 &        0.976 &       0.999 &        - &  18.23 &  $0.249\pm0.070$ &  $-0.218\pm0.108$ \\
      &  4953602734161644672 &  40.462090 & -36.953337 &        0.927 &       0.999 &        - &  18.94 &  $0.395\pm0.141$ &  $-0.439\pm0.197$ \\
      &  5063569255744340992 &  37.411716 & -32.515749 &        0.866 &       1.000 &        - &  18.28 &  $0.278\pm0.083$ &  $-0.254\pm0.128$ \\
      &  5049970972669630464 &  40.059478 & -35.905110 &        0.794 &       0.982 &        - &  18.42 &  $0.587\pm0.097$ &  $-0.153\pm0.137$ \\

\hline
\end{tabular}
\end{table*}

\tabcolsep=0.12cm
\begin{table*}
\centering
\caption{Continued: possible bright ($G \leq 19$ mag) member candidates beyond the King limiting radius for Carina.}
\label{tab:Member_C3}	
\begin{tabular}{lcccccccrr} 
\hline
Galaxy & Source ID & $\alpha$ & $\delta$ &  $P_{Space+PM}$ &  $P_{PM}$ & $V_{r}$ &   $G$ &  $\mu_{\alpha}cos\delta$ &     $\mu_{\delta}$ \\
 
&  & [deg] & [deg] & & & [${\rm km~s^{-1}}$] &   [mag] & [${\rm mas~yr^{-1}}$] & [${\rm mas~yr^{-1}}$] \\
\hline
Carina     &  5502112320339516672 &  101.801148 & -50.858992 &        0.902 &       0.962 &        237.7 &  17.23 &  $0.430\pm0.078$ &   $0.208\pm0.085$ \\
      &  5502484603810420096 &   98.874167 & -51.559791 &        0.848 &       0.966 &        - &  17.60 &  $0.460\pm0.096$ &   $0.205\pm0.095$ \\
      &  5502134271915130752 &  102.268094 & -51.070889 &        0.823 &       0.982 &        - &  17.43 &  $0.510\pm0.084$ &   $0.195\pm0.085$ \\
      &  5502698046503172608 &   99.017039 & -51.051548 &        0.778 &       0.901 &        - &  18.74 &  $0.687\pm0.173$ &   $0.258\pm0.160$ \\
      &  5501694707782516992 &   99.904891 & -51.692459 &        0.757 &       0.909 &        - &  18.71 &  $0.673\pm0.166$ &   $0.226\pm0.188$ \\
      &  5502715157652966912 &   98.843143 & -50.915351 &        0.699 &       0.925 &        - &  18.37 &  $0.492\pm0.139$ &   $0.275\pm0.144$ \\
      &  5503082845509707264 &  102.633411 & -49.614035 &        0.661 &       0.991 &        - &  16.81 &  $0.558\pm0.057$ &   $0.095\pm0.059$ \\
      &  5502003292595409792 &  101.300724 & -51.628797 &        0.619 &       0.942 &        - &  18.35 &  $0.419\pm0.143$ &   $0.071\pm0.143$ \\
      &  5502783086856009472 &   98.546699 & -50.644352 &        0.361 &       0.910 &        - &  18.45 &  $0.390\pm0.143$ &   $0.106\pm0.190$ \\
      &  5503162968122548864 &  101.382482 & -49.778490 &        0.334 &       0.916 &        - &  18.65 &  $0.619\pm0.150$ &   $0.276\pm0.178$ \\
      &  5502321601210343552 &   98.433843 & -52.066497 &        0.284 &       0.914 &        - &  18.77 &  $0.667\pm0.160$ &  $-0.039\pm0.179$ \\
      &  5502693201779934336 &   98.332298 & -50.618324 &        0.255 &       0.903 &        - &  18.63 &  $0.423\pm0.179$ &   $0.202\pm0.215$ \\
      &  5550731689430964864 &   98.113615 & -50.601210 &        0.251 &       0.928 &        - &  18.31 &  $0.678\pm0.147$ &   $0.005\pm0.154$ \\
      &  5507927225023057920 &  104.068619 & -50.782767 &        0.146 &       0.957 &        - &  17.75 &  $0.538\pm0.129$ &   $0.049\pm0.115$ \\
      &  5551333431529187712 &  100.284868 & -48.954146 &        0.135 &       0.974 &        - &  18.12 &  $0.553\pm0.116$ &   $0.089\pm0.114$ \\
      &  5551161568414174592 &   99.067310 & -49.096048 &        0.128 &       0.980 &        - &  17.24 &  $0.622\pm0.068$ &   $0.062\pm0.074$ \\
      &  5503091538523737856 &  102.962540 & -49.427400 &        0.118 &       0.925 &        - &  18.27 &  $0.402\pm0.141$ &   $0.019\pm0.149$ \\
      &  5498748841255266304 &  102.372292 & -52.200728 &        0.094 &       0.940 &        - &  18.01 &  $0.408\pm0.124$ &   $0.043\pm0.135$ \\
      &  5503316968470246016 &  102.418646 & -48.895997 &        0.092 &       0.953 &        - &  18.15 &  $0.582\pm0.127$ &   $0.199\pm0.161$ \\
      &  5501115746195240960 &   96.526764 & -51.811424 &        0.087 &       0.927 &        - &  18.48 &  $0.555\pm0.150$ &   $0.266\pm0.162$ \\
      &  5500861621573059200 &   96.769550 & -52.268656 &        0.085 &       0.927 &        - &  18.49 &  $0.515\pm0.173$ &  $-0.029\pm0.156$ \\
      &  5508291227794323456 &  104.291538 & -49.549433 &        0.082 &       0.946 &        - &  18.13 &  $0.619\pm0.126$ &   $0.168\pm0.157$ \\
      &  5508226567063203456 &  104.582542 & -49.788463 &        0.079 &       0.949 &        - &  17.11 &  $0.411\pm0.070$ &   $0.224\pm0.107$ \\
      &  5508223199808847616 &  104.542154 & -49.855331 &        0.070 &       0.938 &        - &  18.15 &  $0.492\pm0.132$ &   $0.281\pm0.194$ \\
      &  5501047301592401792 &   96.267243 & -52.214214 &        0.065 &       0.932 &        - &  18.55 &  $0.509\pm0.216$ &   $0.062\pm0.166$ \\
      &  5498030654002855168 &  100.777890 & -53.922374 &        0.060 &       0.987 &        - &  16.61 &  $0.486\pm0.064$ &   $0.111\pm0.056$ \\
      &  5498842162304146432 &  103.086474 & -51.882643 &        0.056 &       0.907 &        - &  18.78 &  $0.517\pm0.237$ &   $0.155\pm0.227$ \\
      &  5509107065421158912 &  103.469718 & -48.969735 &        0.053 &       0.929 &        - &  18.70 &  $0.490\pm0.166$ &   $0.262\pm0.173$ \\
      &  5501709894786503040 &  100.809104 & -52.794495 &        0.053 &       0.913 &        - &  18.58 &  $0.501\pm0.165$ &   $0.310\pm0.191$ \\
      &  5551317114950058496 &  101.100594 & -48.724560 &        0.050 &       0.940 &        - &  17.97 &  $0.484\pm0.115$ &  $-0.045\pm0.143$ \\
      &  5498321784065583232 &  100.374342 & -53.109248 &        0.042 &       0.929 &        - &  18.43 &  $0.548\pm0.205$ &   $0.121\pm0.194$ \\
      &  5498583781367476480 &  103.555969 & -52.610894 &        0.038 &       0.963 &        - &  17.57 &  $0.401\pm0.092$ &   $0.174\pm0.098$ \\
      &  5504916757888065152 &  104.810045 & -51.081224 &        0.037 &       0.929 &        - &  18.49 &  $0.555\pm0.189$ &   $0.226\pm0.196$ \\
      &  5503327997946913792 &  102.974485 & -48.588524 &        0.035 &       0.929 &        - &  18.47 &  $0.621\pm0.196$ &   $0.177\pm0.164$ \\
      &  5501271945562728576 &   98.379568 & -53.238275 &        0.030 &       0.903 &        - &  17.97 &  $0.594\pm0.130$ &  $-0.077\pm0.143$ \\
      &  5504817084584768256 &  104.635657 & -51.654347 &        0.026 &       0.927 &        - &  17.94 &  $0.387\pm0.110$ &   $0.033\pm0.113$ \\
      &  5498252377393951488 &  100.410125 & -53.444287 &        0.017 &       0.902 &        - &  18.78 &  $0.457\pm0.187$ &   $0.296\pm0.202$ \\
      &  5551187024683306624 &   98.461590 & -49.023626 &        0.017 &       0.901 &        - &  18.58 &  $0.726\pm0.152$ &   $0.203\pm0.151$ \\
      &  5498511144880789632 &  103.921354 & -53.008845 &        0.010 &       0.928 &        - &  18.04 &  $0.678\pm0.119$ &   $0.158\pm0.130$ \\

\hline
\end{tabular}
\end{table*}


\bsp	
\label{lastpage}
\end{document}